\documentclass[manuscript,nonacm]{acmart}

\usepackage{tabularx}
\usepackage{microtype}
\usepackage{fancyhdr}
\AtBeginDocument{%
  \fancypagestyle{firstpagestyle}{%
    \fancyhf{}%
    \fancyfoot[RO,LE]{\small\thepage}%
    \fancyfoot[RE,LO]{\footnotesize Author version (2022-01-14).}%
  }%
}
\usepackage{color, colortbl}
\usepackage{graphicx}
\usepackage{subcaption}
\usepackage{enumitem}

\newcolumntype{B}{X}
\newcolumntype{s}{>{\hsize=.20\hsize\centering\arraybackslash}X}
\newcolumntype{M}{>{\hsize=.37\hsize}X}

\definecolor{Gray}{gray}{0.9}

\newcounter{p}
\newcommand{\perspective}[2]{\noindent\fcolorbox{darkgray}{lightgray!20}{%
    \minipage[t]{\dimexpr\linewidth-2\fboxsep-2\fboxrule\relax}
        \textbf{#1}\\
        #2
    \endminipage}}
\begin{document}

\title{Cognition in Software Engineering: A Taxonomy and Survey of a Half-Century of Research}

\author{Fabian Fagerholm}
\orcid{0000-0002-7298-3021}
\affiliation{%
  \institution{Aalto University}
  \country{Finland}
}
\affiliation{%
  \institution{Blekinge Institute of Technology}
  \country{Sweden}}
\email{fabian.fagerholm@aalto.fi}
\author{Michael Felderer}
\orcid{0000-0003-3818-4442}
\affiliation{%
  \institution{University of Innsbruck}
  \country{Austria}
}
\affiliation{%
  \institution{Blekinge Institute of Technology}
  \country{Sweden}
}
\email{michael.felderer@uibk.ac.at}
\author{Davide Fucci}
\orcid{0000-0002-0679-4361}
\affiliation{%
  \institution{Blekinge Institute of Technology}
  \country{Sweden}
}
\email{davide.fucci@bth.se}
\author{Michael Unterkalmsteiner}
\orcid{0000-0003-4118-0952}
\affiliation{%
  \institution{Blekinge Institute of Technology}
  \country{Sweden}
}
\email{michael.unterkalmsteiner@bth.se}
\author{Bogdan Marculescu}
\orcid{0000-0002-1393-4123}
\affiliation{%
  \institution{Kristiania University College}
  \country{Norway}
}
\email{bogdan.marculescu@kristiania.no}
\author{Markus Martini}
\orcid{0000-0003-2637-4804}
\affiliation{%
  \institution{University of Innsbruck}
  \country{Austria}
}
\email{markus.martini@uibk.ac.at}
\author{Lars G\"oran Wallgren Tengberg}
\orcid{0000-0002-1826-0148}
\affiliation{%
  \institution{University of Gothenburg}
  \country{Sweden}
}
\email{larsgoran.wallgren@psy.gu.se}
\author{Robert Feldt}
\orcid{0000-0002-5179-4205}
\affiliation{%
  \institution{Chalmers University of Technology}
  \country{Sweden}
}
\affiliation{%
  \institution{Blekinge Institute of Technology}
  \country{Sweden}
}
\email{robert.feldt@chalmers.se}
\author{Bettina Lehtelä}
\orcid{0000-0002-2814-4386}
\affiliation{
  \institution{Aalto University}
  \country{Finland}
}
\email{bettina.lehtela@aalto.fi}
\author{Bal\'{a}zs Nagyv\'{a}radi}
\orcid{0000-0003-4835-6738}
\affiliation{
  \institution{Aalto University}
  \country{Finland}
}
\email{balazs.nagyvaradi@aalto.fi}
\author{Jehan Khattak}
\orcid{0000-0002-4912-1757}
\affiliation{
  \institution{Aalto University}
  \country{Finland}
}
\email{jehan.khattak@aalto.fi}

\begin{abstract}
Cognition plays a fundamental role in most software engineering activities.
This article provides a taxonomy of cognitive concepts and a survey of the literature since the beginning of the Software Engineering discipline.
The taxonomy comprises the top-level concepts of perception, attention, memory, cognitive load, reasoning, cognitive biases, knowledge, social cognition, cognitive control, and errors, and procedures to assess them both qualitatively and quantitatively.
The taxonomy provides a useful tool to filter existing studies, classify new studies, and support researchers in getting familiar with a (sub) area.
In the literature survey, we systematically collected and analysed 311 scientific papers spanning five decades and classified them using the cognitive concepts from the taxonomy.
Our analysis shows that the most developed areas of research correspond to the four life-cycle stages, software requirements, design, construction, and maintenance. 
Most research is quantitative and focuses on knowledge, cognitive load, memory, and reasoning. 
Overall, the state of the art appears fragmented when viewed from the perspective of cognition.
There is a lack of use of cognitive concepts that would represent a coherent picture of the cognitive processes active in specific tasks.
Accordingly, we discuss the research gap in each cognitive concept and provide recommendations for future research.
\end{abstract}



\maketitle
\renewcommand{\shortauthors}{Fagerholm et al.}

\section{Introduction}
Cognition is the \emph{``collection of mental processes and activities used in perceiving, remembering, thinking, and understanding, and the act of using those processes''}~\citep{ashcraft2002}. These processes include the formation of knowledge, creation and use of memory, problem solving, decision making, attention, and comprehension. Studying the quality of mental processes and how they can be affected is of general interest in psychology, but is also relevant for knowledge-intensive fields such as Software Engineering (SE).
From an SE perspective, it is interesting to describe and measure (quantitatively and qualitatively) cognitive concepts, since this enables the systematic evaluation of interventions aimed at improving mental processes related to an engineering task, and subsequently the performance in the task.

Researchers have studied behavioural aspects of SE since the dawn of the discipline~\citep{curtis1984fifteen}.
The question of how the human mind processes information related to software development activities has been of interest of SE research~\citep{cunha_decision-making_2016,lenberg2015, parviainen_knowledge-related_2014,stacy_cognitive_1995}.
Understanding cognitive processes in a quantifiable manner is a powerful lever when engineering effective processes and tools that support software development activities.
Therefore, to advance SE research and understand which cognitive concepts have seen attention in combination with particular software development activities, a systematic categorisation of SE research using a taxonomy of cognitive concepts is valuable.
While there are taxonomies that cover aspects of cognition in SE (e.g., human errors~\cite{ANU2018112}, cognitive biases~\cite{mohanani2018cognitive}, cognitive load~\cite{gonccales2019measuring}), there is not, to the best of our knowledge, taxonomy covering the wide spectrum of concepts that have been studied in the general field of psychology.  

This article provides \textit{i)} an overview of the vast and fragmented research on cognition in SE,  \textit{ii)} definitions of central cognitive concepts, processes, and activities, and \textit{iii)} approaches to assess them. We propose a taxonomy of cognitive concepts and assessment procedures relevant for SE. We then apply this taxonomy to characterise SE research focusing on cognitive aspects---with articles published between 1973\footnote{This date does not indicate a particular event but the publication date of the earliest paper we classified.} and 2020.

The contribution of this study is threefold, providing:
\begin{enumerate}
\item A taxonomy of cognitive concepts and a framework to measure and assess these concepts, developed in collaboration with experts from cognitive and organisational psychology.
\item An overview of the research in SE, characterised by which cognitive aspects were studied and how they were measured and assessed.
\item A discussion on which knowledge gaps exist with regard to cognitive aspects of SE.
\end{enumerate}

This paper focuses on cognitive concepts in professional software development.
We do not take into account aspects dedicated to \textit{i)} pedagogical, educational, and learning activities~\cite{dybaa2014reflective} as well as \textit{ii)} knowledge management and software design decision ~\cite{capilla201610,razavian2016reflective}, which also have cognitive aspects.

The remainder of this paper is structured as follows.
In Section~\ref{sec:relatedwork}, we provide definitions of cognition, present prior reviews of the literature related to cognition in SE, argue for the need for the present review, discuss taxonomies in SE, and set out our objectives related to the taxonomy. In Section~\ref{sec:method}, we describe the research approach and methods used in this article. In particular, we describe the taxonomy development and literature search, selection, and analysis, and discuss the threats to validity of our method choices. In Section~\ref{sec:taxonomy}, we present the taxonomy of cognitive concepts and assessment procedures. We describe the application of the taxonomy and the results of the literature study in Section~\ref{sec:results}. In Section~\ref{sec:discussion}, we discuss the study results and conclude the paper in Section~\ref{sec:conclusion}.

\section{Preliminaries and Related Work}
\label{sec:relatedwork}

From technical details to the broadest perspectives on design and requirements, making software involves people thinking. Before software runs on our computing devices, it is imagined in the minds of software professionals---those involved in product design, project management, requirements, architecture, user experience, testing, programming, and other development and design activities.

Researchers have investigated thinking among software professionals for decades (e.g.,~\cite{weinberg1971,shneiderman1976,soloway1982,letovsky1986,detienne1990,detienne_assessing_1997}), providing guidance, methods, and tools for those involved in the development endeavour. Some early studies concerned programming---reading and understanding source code---with implications for variable naming, programming language design, finding, and preventing programming errors. In the past decades, various topics related to human cognition in SE have been investigated.

In this section, we first discuss cognition as a psychological concept to demarcate the focus of this article. We then provide an overview of past work that, similarly to our aim, reviewed studies that pay particular attention to cognitive aspects of SE. We first look at historic reviews that suggest that cognitive aspects were studied from the inception of the SE field. Then, we summarise the findings of contemporary reviews in SE in general and in specialised sub-fields. Accordingly, we provide a motivation as to why our review is necessary and how it contributes to the existing body of knowledge. Finally, we discuss the concept of a taxonomy, how taxonomies are used in SE, and outline the goals of the taxonomy presented in this article.

\subsection{What is Cognition?}
\label{sec:what-is-cognition}

\emph{Cognition} is an umbrella term for a host of different conscious and unconscious mental processes. It refers to the acquisition and processing of information~\cite{colman2009}, and spans processes of perception, attention, memory, thinking, and problem-solving~\cite{ashcraft2002}. For example, during programming, we generate, maintain and manipulate a representation of source code in our working memory based on our (expert) knowledge of different programming languages which is stored in our long-term memory. We can direct our focus of attention to specific parts of the code or think about and mentally simulate a solution to a problem. Programming integrates multiple, dynamic, and interacting mental processes which can partly run in parallel. The same is true for other software design and development tasks, such as working with requirements, architectures, and testing, dealing with project management, providing effort estimates, and executing specific development processes and techniques.

Cognition can be conceptually separated from emotion and motivation, although they arise from interacting neural circuits and processes~\cite{pessoa2008,salzman2010}. The human cognitive system and cognitive behaviour, such as decision-making, is also affected by whether an activity occurs for a single individual or in a group~\cite{cacioppo2011}. Furthermore, cognition can also be conceptualised from a sociological perspective that considers interactions between humans and with culturally significant objects to constitute cognitive processes. This is the case in distributed cognition theory, according to which cognition is not only individuaĺ, but is embedded and situated in external artefacts, groups, and cultural systems for interpreting reality~\citep{hutchins1995}.

In this article, we focus on individual information processing and how it has been studied in SE. We have chosen to put less emphasis on emotion and motivation, and to include only a small subset of cognition in social situations in which the focus is mainly social-psychological and concentrates on the individual. Prior reviews have given overviews of the research on motivation~\cite{beecham2008,franca2011} and emotion~\cite{sanchezgordon2019} in SE.

Cognition can be studied using various means, ranging from psycho-physiological measurements to open observations of behaviour. Different assessment procedures serve different purposes---from detailed information on, for example, perceptual processes, to uses as part of process improvement or as input to software development tools. We have chosen to include all kinds of assessment procedures that we have found in the literature, to illustrate the wide variety of available approaches.

\subsection{Historic reviews}

In the early 1980s, Sheil~\cite{sheil_psychological_1981} reviewed work on the psychology of programming published since the late 1960s, when the field of SE emerged as an independent research area.  The review focuses on programming notation (conditionals, control flow, data types), practices (flow-charting, indenting, variable naming, commenting) and tasks (learning to program, coding, debugging), and their impact on developer performance. Sheil criticises the methodological flaws in how these phenomena have been studied, in particular the design of experimental treatments, practice effects (learning), failure to compensate for individual variability, and not considering effect size in statistical tests.

Curtis~\cite{curtis1984fifteen} follows in 1984 by reviewing studies on individual differences of programmers in terms of expertise, problem solving, fault detection, and learning to program. He points to the lack of research, in the reviewed time-period from 1968 to 1984, on problem-solving processes in requirements engineering and software design. With the popularisation of object-oriented programming in the 1980s and 1990s, more empirical studies on software design and its consequences on cognition were performed, as reviewed by D\'etienne~\cite{detienne_assessing_1997}.

These early reviews culminate in 1986 in a call for collaboration between psychologists and computer scientists~\cite{curtis_software_1986}. In this review, Curtis \textit{et al}.\ summarise studies on problem solving in unstructured domains, use of programming languages, knowledge representation for programmers, variable naming, and research on the parallels between text and source code understanding. Furthermore, the review discusses methodological approaches (e.g., experiments) and issues in designing and conducting them, and points towards the need for research in conceptual and mental modes, psychological factors affecting programming, transfer of skills, communication media between programmers, and problem solving representations.

\subsection{Contemporary reviews}

Here, we discuss reviews pertinent to cognitive aspects that have been published between 2014 and 2019. We begin by looking at four reviews that studied the SE discipline as a whole and two reviews that focused on software project management.

Lenberg et al.~\cite{lenberg2015} conducted a systematic literature review (SLR) with the purpose of identifying concepts relevant for the proposed research area of Behavioural SE (BSE). They identified 55 concepts, cognition being one of them, within the area of BSE. Between 1997 and 2013, the most studied BSE concepts were communication, personality, and job satisfaction.

Personality in SE research was the focus of a systematic mapping study by Cruz et al.~\cite{cruz_forty_2015}. They found that, in the reviewed time-frame from 1970 until 2010, personality was studied mostly in the context of pair-programming, but also in education, team effectiveness, software-process allocation, software engineer personality characteristics, individual performance, team process, behaviour and preferences, and leadership performance. They point out that even after 40 years of research, the field does not yet seem mature. Conflicting evidence indicates that the results presented in the studies cannot be directly applied, as they may not produce the desired results.

Sharafi et al.~\cite{sharafi_systematic_2015} conducted an SLR of studies utilising eye-tracking to record visual attention as a proxy for cognitive processes. They found 36 studies investigating the engineers performing tasks pertinent to model comprehension, code comprehension, debugging, collaborative interactions (e.g., studying eye-movement as cues in collaborative tasks), and traceability recovery.

Blackwell et al.~\cite{blackwell_fifty_2019} reviewed the literature pertaining to the psychology of programming, published in the Journal of Man-Machine Studies and in the Journal of Human-Computer Studies between 1969 and 2019. They describe the evolution of the psychology of programming field during the five decades and observe a change in research focus: in the early days, from individual developers and small scale programs (1970s), through large code bases and programs where professional skills and processes become more important (1980s and 1990s), to contemporary issues where software development is a social activity that is conducted in different contexts (2000s and 2010s). 

Advances and novel research directions in psychology often find their way into SE research and suggest new ways of looking at different phenomena in the field. For example, the research program on cognitive biases received mainstream attention in 2002 with a Prize in Economic Sciences in Memory of Alfred Nobel (c.f., Kahneman~\cite{kahneman2011}), likely influencing research in SE as well. In 2014, Shepperd~\cite{shepperd_cost_2014} discusses how cost estimations are affected by the planning fallacy, a preference for case-specific over distributional evidence, the peak-end rule, and the anchoring effect. In 2016, Cunha et al.~\cite{cunha_decision-making_2016} briefly review how cognitive biases have been studied in the context of decision-making in software project management.

\subsection{Need for a review on cognitive concepts in SE}
Looking at the reviews discussed above, both historic and contemporary ones, we observe that cognition has either (1) not been in the focus of the investigation (e.g.,  behavioural SE by Lenberg et al.~\cite{lenberg2015} or personality by Cruz~\cite{cruz_forty_2015}), (2) was studied in a narrow area of SE, such as software construction, design and testing (e.g., Blackwell et al.~\cite{blackwell_fifty_2019}, which is also limited to two journals, or Sheil~\cite{sheil_psychological_1981} and Curtis et al.~\cite{curtis1984fifteen,curtis_software_1986}, which were also done more than 30 years ago) or software project management~\cite{shepperd_cost_2014,cunha_decision-making_2016}, or (3) focused on a particular method to study cognitive processes (e.g., eye-tracking by Sharafi et al.~\cite{sharafi_systematic_2015}).

Reviews of cognition in SE have the double challenge of capturing both the breadth of topics in SE, and the breadth of cognition as a psychological concept. Curtis~\cite{curtis1984fifteen} observed already 25 years ago that the research on behavioural aspects of SE has not been organised into a sub-field of SE or psychology. He attributes this to the difficulty of integrating the mixture of paradigms borrowed from psychology.  While the SE Body of Knowledge (SWEBOK)~\citep{swebok} recognises group dynamics and psychology as an aspect of SE professional practice, topics such as individual cognition and problem solving are merely touched upon. Existing taxonomies in SE do not comprehensively cover people issues in general~\cite{usman2017}, and thus neither cognition in particular.

To the best of our knowledge, no comprehensive taxonomy of cognitive concepts relevant to SE has yet been proposed and used to characterise the field. With the taxonomy and review presented in this paper, we fill this gap and complement previous reviews.

\subsection{Taxonomies in SE}
\label{sub:taxonomies}

Classifications in general, and taxonomies in particular, provide a common terminology for knowledge sharing, a better understanding of the relationships between concepts in particular fields, and support in decision making and identifying gaps in knowledge~\cite{usman2017}. A key benefit of taxonomies, and of classifications in general, is to ``increase cognitive efficiency and facilitate inferences,'' as well as facilitating education~\cite{ralph2019}. An example of the application of a taxonomy to SE research is the SE Body of Knowledge (SWEBOK)~\cite{swebok} that provides an overview of the knowledge areas that are relevant to the field.

Taxonomies provide researchers with the ability to reason about more generally-applicable classes of a domain, rather than restricting them to individual instances~\cite{ralph2019}. By grouping concepts and their relationships based on some assessment of similarity, researchers can establish useful inferences, including counter-intuitive causal or temporal relationships based on data. Ralph~\cite{ralph2019} discusses dimensions for analysing and assessing taxonomies: (1) the degree to which the class structure reflects similarities and dissimilarities observed between actual instances, (2) the ability to draw useful conclusions about instances based on their membership in relevant classes, and (3) the degree to which the taxonomy meets its goal.

Ralph also discusses the philosophical grounding for taxonomies and their use~\cite{ralph2019}. The author notes that teleological, dialectic, and life-cycle process theories include taxonomies and are most suited to being described by such classifications. In addition, the paper provides examples of the cognitive efficiency gained by using taxonomies, and how it can be further used to draw more general conclusions from the available data~\cite{ralph2019}.

We define a taxonomy of psychological concepts used in SE in accordance with the guidelines provided in Usman et al.~\cite{usman2017}.
In particular, we focus on the definition of a common terminology to improve research communication and education, and on identifying knowledge gaps (c.f.~\cite{ralph2019}). We first define the taxonomy based on domain knowledge from SE and grounded in psychology literature. Then, we use it as a framework to investigate the extent to which cognitive concepts have been researched in SE, and to identify gaps in the current knowledge.

\section{Research Methodology}
\label{sec:method}

Based on the literature, we observed that a rather large body of work exists at the overlap between SE and psychology, in particular the study of human cognition and SE activities. Accordingly, we set out to systematically create a taxonomy of cognitive concepts and assessment procedures, and to classify the existing scientific literature to understand development, gaps, and future directions of the field.

Accordingly, this study is driven by the following research questions:

\begin{itemize}
\item[RQ1:] How can cognitive concepts relevant for SE be classified and assessed?
\item[RQ2:] What is the state of the art of cognitive concepts in SE?
\begin{itemize}
    \item[RQ2.1:] In which context have cognitive concepts been studied in SE?
    \item[RQ2.2:] How has the study of cognitive concepts in SE developed over time?
    \item[RQ2.3:] Which SE knowledge areas have been studied under the lens of cognitive concepts?
    \item[RQ2.4:] How have cognitive concepts in SE been assessed?
    \item[RQ2.5:] Which theories have been used to study cognitive concepts in SE?
\end{itemize}
\item[RQ3:] What gaps and potential future directions exist in the field of cognition in SE?
\end{itemize}

We assembled an interdisciplinary team of researchers to work on answering the research questions. Table~\ref{tab:researchers} lists the researchers, their main discipline and the tasks they were involved in this research. The choice of tasks was first and foremost driven by expertise, personal preference and, lastly, time availability. We refer to the researchers' acronyms when describing the tasks later in this section. We conducted the research in two phases. During the 1\textsuperscript{st} phase, we executed the initial search that included primary studies until 2017. We conducted the 2\textsuperscript{nd} phase in 2020 to include primary studies published after 2017.

\begin{table}
  \centering
  \footnotesize
  \caption{Researchers involved in this work, their main disciplines, and tasks.}
  \label{tab:researchers}
  \begin{tabular}{p{0.22\textwidth} p{0.08\textwidth} p{0.17\textwidth} p{0.43\textwidth}}
    \toprule
    \multicolumn{4}{c}{1\textsuperscript{st} Phase} \\
    \midrule
    Name & Acronym & Main discipline & Tasks \\
    \midrule
    Fabian Fagerholm & FFA & SE & taxonomy construction, paper selection, data extraction, validation  \\
    Michael Felderer & MFE & SE & taxonomy construction, paper selection, data extraction, validation \\
    Robert Feldt & RFE & SE & paper selection, data extraction, validation \\
    Davide Fucci & DFU & SE & validation \\
    Michael Unterkalmsteiner & MUN & SE & paper selection, data extraction, validation, classification \\
    Bogdan Marculescu & BMA & SE & data extraction, validation \\
    Markus Martini & MMA & Psychology & taxonomy construction \\
    Lars-Göran Wallgren Tengberg & LTE & Psychology & taxonomy construction \\
    \midrule
    \multicolumn{4}{c}{2\textsuperscript{nd} Phase} \\
    \midrule
    Name & Acronym & Main discipline & Tasks \\
    \midrule
    Fabian Fagerholm & FFA & SE & paper selection, data extraction, validation, classification \\
    Bettina Lehtel\"{a} & BLE & SE & paper selection, data extraction, validation, classification \\
    Bal\'{a}zs Nagyv\'{a}radi & BNA & SE & paper selection, data extraction, validation, classification \\
    Jehan Khattak & JKH & Human-Computer Interaction & paper selection, data extraction, validation, classification \\
    
    \bottomrule
  \end{tabular}
\end{table}

\begin{table}
    \centering
    \footnotesize
    \caption{Included and excluded primary studies during the activities of the1\textsuperscript{st} and 2\textsuperscript{nd} phase of the literature survey}
    \label{tab:incexlpapers}
    \begin{tabular}{lllll}
        \toprule
         & \multicolumn{2}{c}{1\textsuperscript{st} Phase} & \multicolumn{2}{c}{2\textsuperscript{nd} Phase} \\
         Activity & Included & Excluded & Included & Excluded \\
         \midrule
         Keyword search & 738 & - & 1030 & - \\
         Selection & 208 & 530 & 96 & 934 \\
         Data extraction & 129 & 79 & 96 & 0 \\
         Backward snowball sampling & 113 & 0 & - & - \\
         Classification & 217 & 25 & 96 & 0 \\
         \midrule
         Total included primary studies & \multicolumn{4}{r}{311} \\
         \bottomrule
    \end{tabular}
\end{table}

To answer RQ1, we invited experts in cognitive and behavioural psychology to support the taxonomy development, (see  Section~\ref{sec:taxdev}).
To answer RQ2, we conducted a literature survey driven by keyword search and backward snowball sampling~\cite{wohlin_guidelines_2014} as described in Sections~\ref{sec:search}--\ref{sec:updatedsearch}.
The quantitative results of included and excluded primary studies are summarised in Table~\ref{tab:incexlpapers}. Finally, to answer RQ3, we classified the main research contributions of the identified relevant studies using the developed taxonomy on cognitive concepts and assessment procedures, and the knowledge areas presented in SWEBOK, as described in Section~\ref{sec:synthesis}. Threats to the validity of our research methodology are discussed in Section~\ref{sec:threats}.

\subsection{Taxonomy development: Cognitive concepts and assessment procedures}\label{sec:taxdev}
To classify existing and future work on cognitive research in SE, FFA and MFE developed, together with two experts (MMA, LTE), a taxonomy of cognitive concepts and assessment procedures. Based on searches of the literature, and according to the best of our knowledge as well as the expertise of the involved psychologists, no suitable taxonomy of cognitive concepts existed. Taxonomies do exist in specific areas of psychology, such as personality and developmental psychology, as well as certain specific topical areas within organisational and cognitive psychology. However, these are too specific to cover the range of cognitive concepts investigated in SE. Since we do not aim to create a taxonomy for use in psychology, we decided to develop a new taxonomy for SE, based on concepts from psychology. The steps of the taxonomy development process are summarised in Table~\ref{tab:taxdev} and explained next.

In step 1, the two experts (MMA, LTE) independently created two taxonomies each, one for cognitive concepts and one for cognitive assessment procedures. They used their knowledge of psychological research and practice, textbooks, and seminal works to create potential topic areas for the taxonomy. To make the task more specific and concrete, we established constraints during the development of the taxonomy and developed concrete exemplary concepts and assessment procedures.

\begin{table}
    \centering
    \footnotesize
    \caption{The taxonomy development, refinement and validation process}
    \label{tab:taxdev}
    \begin{tabular}{llp{7.5cm}l}
         \toprule
         Step & Name & Description & Output \\
         \midrule
         1 & Initial development & MMA and LTE independently created candidate taxonomies. & Two candidate taxonomies \\
         2 & Initial testing & MMA, LTE applied the candidate taxonomies on sample studies (from the ongoing literature survey) and refined structure and content. & Refined two candidate taxonomies\\
         3 & Integration & FFA and MFE integrated the candidate taxonomies. & One integrated taxonomy \\
         4 & Review & MMA and LTE reviewed the integrated taxonomy. & Reviewed integrated taxonomy \\
         5 & Validation & MUN, FFA, BLE, BNA and JKH applied the taxonomy on 311 papers originating from the literature review & Validated taxonomy\\
         \bottomrule
    \end{tabular}
\end{table}

In the \emph{constraints}, we stated that the purpose of the taxonomies is to classify papers that concern cognitive aspects in SE. Further, we established that the taxonomies should:
\begin{enumerate}
\item Have distinct categories with minimal overlap so that papers can be unambiguously categorised.
\item Have explanations for each category, permitting categorisation as objectively as possible.
\item Capture research areas and assessment procedures commonly used in the field of cognitive psychology.
\item Have practical utility in supporting an understanding of which areas are covered and which are not, given a set of papers.
\end{enumerate}

In step 2, we provided the experts with \emph{sample papers} and the hint that these papers should be classifiable based on the created taxonomy.
The sample papers originated from the ongoing literature survey described in Section~\ref{sec:search}--\ref{sec:updatedsearch}.
While the full validation was conducted in Step 6, the purpose of this step was to detect any errors of omission in the taxonomy at an early stage. 

We provided \emph{examples} like the following:
\begin{small}
\begin{verbatim}
Given a paper that investigates the cognitive effort of reading program source code,
we have extracted the following information.
  Cognitive Concepts: Cognitive Effort
  Cognitive Measures: Eye gaze
  Assessment: Automatic measurement of eye movement with tool iTrace

This paper could be classified as:
  Concept: Working Memory -> Cognitive Load
  Assessment procedure: Physiological Measures -> Eye Gaze

However, this is just an example and other classifications could be possible as well.
The task is to work out two schemes of classification (one for concepts and another
for assessment procedures) that allow this kind of classification to be performed.
\end{verbatim}
\end{small}

In step 3, FFA and MFE integrated the taxonomies received from each psychologist into one taxonomy of cognitive concepts and assessment procedures.
The integration was done with the goal of achieving the constraints listed above.
Concepts proposed by both experts were retained as such but their position in the taxonomy was chosen considering all other concepts, so as to make the taxonomy logical and parsimonious.
Concepts proposed by only one expert were first included but could later be merged with another concept if too similar, made into a sub-concept, or removed. We initially included emotion- and motivation-related concepts but later pruned them from the taxonomy since, as discussed in Section~\ref{sec:what-is-cognition} our emphasis is on individual information processing, and the SE field has investigated emotion and motivation as separate constructs.
Proposed concepts related but not the same were analysed using the definitions in the reference literature to determine their relationships, and a corresponding hierarchy was created.
There were no concepts where the experts would have proposed differing terminology for essentially the same concept.
There were also no directly conflicting concepts.
During the merging process, we asked the psychologists for further details and supporting references in case of ambiguity or lack of clarity.

The resulting taxonomy consists of two dimensions (cognitive concepts and assessment procedures), top-level categories and sub-categories. For each category and sub-category, FFA and MFE identified definitions originating from the scientific literature in psychology. The dimension of cognitive concepts consists of nine top-level categories and 15 sub-categories while the taxonomy of cognitive assessment procedures consists of two top-level categories and eight sub-categories.

In step 4, the experts reviewed and accepted the integrated taxonomy (shown in Figure~\ref{fig:taxonomy}). 

Finally, in step 5 we classified the set of papers originating from the literature survey on cognitive concepts in SE, which is discussed next.

\subsection{Initial Keyword Search}
\label{sec:search}

To identify SE studies that investigate cognitive concepts, we followed two strategies. First, MUN, MFE and RFE performed a keyword search in six research databases: ACM Digital Library, Emerald Insight, IEEE Xplore, ScienceDirect, Scopus, and Wiley. Then, after paper selection, we performed backward snowball sampling on the identified primary studies, as explained in Section~\ref{sec:snowball}. The search string is composed of Boolean expressions including (1) a set of terms related to cognitive concepts and (2) a set of terms related to knowledge areas from SWEBOK (see Table~\ref{tab:keywords}).

\begin{table}
  \centering
  \footnotesize
  \caption{Terms used in the keyword search.}
  \label{tab:keywords}
  \begin{tabular}{p{0.18\textwidth}p{0.75\textwidth}}
    \toprule
    &  \multicolumn{1}{c}{Terms} \\
    \midrule
    Cognitive concepts &  "cognitive load" OR "cognitive effort" OR "cognitive resource" OR "cognitive cost" OR "cognitive burden" OR "cognitive overload" OR "cognitive style" OR "cognitive capacity" OR "cognitive work" OR "cognitive psychology" OR "cognitive neuroscience" OR "mental capacity" OR "working memory" OR (cognitive AND attention) OR "task difficulty" OR "cognitive fit" OR "cognitive bias"\\
    \multicolumn{1}{c}{AND} & \\
    SE knowledge areas & "software engineering" OR "software development" OR "requirements engineering" OR "software testing" OR "software design" OR "software construction" OR "software process" OR "software maintenance" OR "software configuration" OR "software model" OR "software method" OR "software quality" OR "software economics" OR "software project management" OR "software life cycle"\\
    \bottomrule
  \end{tabular}
\end{table}

We iteratively developed the search terms, starting from \texttt{"cognitive load" AND "software engineering"} by performing test searches and adding new terms to the set of cognitive concepts. We also sampled terms from the key chapters in SWEBOK and added them to the set of SE knowledge areas.

We restricted the search, performed in January 2017, to title, abstract and keywords. After removing duplicate results, proceedings and workshop summaries by inspecting the resulting titles, we collected 738 publications.

\subsection{Selection of relevant papers}
\label{sec:selection}

We defined explicit criteria to determine the relevance of a study. MUN, MFE, and RFE piloted the selection and refined the inclusion and exclusion criteria during this process. The inclusion criterion (IC1) was: \emph{The paper describes or aims at improving an SE activity, using a cognitive concept as a lens.} We complemented this broad definition with very specific exclusion criteria. A paper should be excluded if:
\begin{description}
\item[EC1] The paper is focused on educational or knowledge management aspects and not on SE activities per se.
\item[EC2] The paper is about cognitive aspects of the use of the software/system rather than of its development. The dividing line is whether a study looks at the cognitive aspects of performing the design of a user interface, which is included. A study that looks at the effect of a user interface on the user, considering cognitive aspects, is excluded.
\item[EC3] The paper is about emotional aspects exclusively---i.e., cognitive aspects are not considered.
\item[EC4] The paper is a method paper and the focus is on a method inspired by, or based on, cognitive psychology rather than on application of such a method.
\item[EC5] The paper is a literature review (systematic or ad-hoc) and does not add any new empirical results.
\item[EC6] The cognitive concept is used as a passive motivation, but not actively evaluated or observed in the study.
\item[EC7] The paper is not written in English.
\item[EC8] The paper is not peer-reviewed.
\end{description}

We applied these criteria on the title and abstract of a publication during the selection process, but also referred to them during the extraction (Section~\ref{sec:extraction}) when we read the full-text and needed to exclude a paper based on its (lack of relevant) content or diverging focus that was not clear from reading the abstract alone.

We split the 738 publications resulting from the keyword search into three sets with a 10\% overlap which MUN, MFE and RFE reviewed. Within the 75 publications that were reviewed by all researchers, there was an agreement of 79\% (59) on the inclusion/exclusion decision. In terms of Fleiss' Kappa~\cite{fleiss1971}, the agreement was moderate (0.48). After the entire selection process, we included 208 publications.

\subsection{Data Extraction}\label{sec:extraction}

The data extraction was performed by four researchers (MUN, MFE, RFE, and FFA), using a shared spreadsheet with the data fields shown in Table~\ref{tab:dataextraction}. The meta-data fields were extracted automatically from the search results, while the study setting and cognitive aspects fields required reading the full-text of the paper, together with an interpretation by the respective extractor. The extractor could also decide to exclude a paper, using the exclusion criteria listed in Section~\ref{sec:selection}. This decision was reviewed by at least one other extractor and, in case of disagreement, discussed among all extractors. In this stage, we excluded 79 publications and extracted data from 129.

\begin{table}
  \centering
  \caption{Data extraction form}
  \label{tab:dataextraction}
  \begin{tabularx}{\textwidth}{MsB}
    \toprule
    \multicolumn{1}{c}{\textbf{Data field}} & \multicolumn{1}{c}{\textbf{Data type}} & \multicolumn{1}{c}{\textbf{Value}} \\
    \midrule
    \multicolumn{3}{c}{\textit{Meta-data}} \\
    \midrule
    Publication venue & free text & Name of the outlet in which the paper was published. \\ \rowcolor{Gray}
    Year & free text & Year in which the paper was first published. \\
    \midrule
    \multicolumn{3}{c}{\textit{Study setting}} \\
    \midrule
    Context & restricted & Based on the study participants, the context was determined as academia, industry, or both. If the study did not involve participants, we set the context to academia. \\ \rowcolor{Gray}
    Type & restricted & We adapted Wieringa et al.~\cite{wieringa_requirements_2005} to the following study types: solution proposal, empirical study, experience paper, philosophical paper, opinion paper. Since one publication can report on multiple studies, several study types can be selected for one paper. \\
    \midrule
    \multicolumn{3}{c}{\textit{Cognitive aspects}} \\
    \midrule
    Cognitive concepts & free text & Concepts discussed in relation to SE and cognition. E.g., procedural or structural problem solving.\\ \rowcolor{Gray}
    Cognitive measures & free text & The item in relation to cognition that is measured. E.g., comprehension based on structural or procedural solution.\\
    Assessment & free text & The method by which the cognitive concept is measured. E.g., questions relating to the programming task at hand.\\ \rowcolor{Gray}
    Added value & free text & A summary stating to what extent the cognitive concept is central or tangential to the conducted study reported in the paper. \\
    Theory & free text & Name(s) of theories, models, or frameworks related to cognitive concepts that are mentioned in the paper. \\\rowcolor{Gray}
    Outcome/contribution & free text & A summary or copy of the papers' main contribution. Can be empty. \\
    SWEBOK area & restricted & The sixteen SWEBOK knowledge areas (KAs), with possible multiple selection. A KA was selected if the cognitive concept was studied \emph{in} or had an impact \emph{on} the KA: software requirements, software design, software construction, software testing, software maintenance, software configuration management, SE management, SE process, SE models and methods, software quality, SE professional practice, SE economy, computing foundations, mathematical foundations, engineering foundations. \\ \rowcolor{Gray}
    Excluded & restricted & Binary choice; decision based on the full-text reading of the paper, applying the inclusion/exclusion criteria. \\
    \bottomrule
  \end{tabularx}
\end{table}

\subsection{Backward Snowball Sampling}
\label{sec:snowball}

During the data extraction, we noted that several referenced studies were relevant according to our inclusion criteria, but not found by the keyword search. At this point, we had two options: (1) add new keywords to the original search, or (2) perform snowball sampling to reach relevant papers from the already identified papers. We opted against the first strategy as it would be difficult to decide which keywords should be added and when to stop the addition of new keywords.  We chose therefore to do backward snowball sampling from the 129 already extracted papers by reading first title, and if necessary, the abstract of referenced papers. MUN, MFE, FFA, and BMA carried out the snowball sampling, resulting in additional 113 papers from which we extracted data.

\subsection{Classification}\label{sec:synthesis}

We used the taxonomies on cognitive concepts and assessment procedures, developed as described in Section~\ref{sec:taxdev}, to further classify the data we extracted from the primary studies, as described in Section~\ref{sec:extraction}. MUN used the extracted values from the data fields \emph{cognitive concepts}, \emph{cognitive measures}, \emph{assessment}, and \emph{added value} (see Table~\ref{tab:dataextraction}) to associate each primary study with one or more dimensions of the two taxonomies, using the definitions shown in Figure~\ref{fig:taxonomy}. In this step, we further excluded 25 publications with the following motivations:
\begin{itemize}
\item The discussion of the cognitive concept was not central to the study but rather used as a motivation/justification.
\item The discussion of the cognitive concept was superficial and lacked details to a degree that made classification impossible.
\end{itemize}
The decision to exclude a publication was discussed among MUN, BMA, MFE, and FFA, resulting in a total of 217 publications that were classified in this step.

\subsection{Updated Review}
\label{sec:updatedsearch}

To bring the most recent publications into the result set, we performed a second round of keyword searching, selection, extraction, and classification.
We opted for a new search rather than a forward snowballing from the set obtained in the first round because we wanted to bring in more recent publications while remaining consistent with respect to the kinds of papers found, and strengthening the reliability of the first round results by verifying the searches. 
BLE, BNA, JKH, and FFA first ran the original keyword searches in June 2020 without any publication year restrictions to verify the search protocol.
When comparing the results, we noticed that some databases returned additional results from the period covered by the first search.
We speculate that some databases may have been updated to include additional publications since the first search, or that their search algorithms may have changed.
We analysed the results of each search to determine the shortest date range that would cover all papers not yet found.
The ranges varied from 1979--2020 to 2017--2020, with most being closer to the latter.
After removing duplicates, we collected 1030 new publications, extending our review to cover papers published and indexed up to May 2020.

FFA, BLE, BNA, and JKH then selected papers to include based on the inclusion and exclusion criteria.
One extractor proposed a selection decision which was then verified by another extractor.
In case of disagreements, the decision was resolved through discussion among all extractors.
This procedure meant that all paper extractions were verified in this phase.
As before, we also excluded papers in the extraction phase using the same verification procedure.
After all exclusions, 96 new publications remained, from which FFA, BLE, BNA, and JKH extracted information and classified the publications according to the two taxonomies.
The extraction and classification of approximately 17\% of these papers was checked by having a different person among these four researchers independently verify them.
We resolved conflicts through discussion sessions between the original extractor and the verifier, while FFA resolved any remaining conflicts. 
Slightly less than 5\% of the extracted papers had very minor discrepancies (e.g., formulation of main results of the study).
Therefore, we did not perform further validation on the remaining 83\% of the papers.

We decided not to perform snowball sampling on the 96 new publications, because the first round of searches and snowball sampling would already cover most of the range where citations from the new publications could be found. Also, this second phase was aimed primarily at extending the range of included papers to cover the years 2017--2020, rather than increasing the set of papers overall.

In total, the final set extracted and classified consists of 311 papers (217 in the first phase and 96 in the second phase) published between 1973 and May 2020\footnote{The final result set with the extracted data is available as online supplementary material~\cite{dataset}.}.

\subsection{Threats to Validity}
\label{sec:threats}

In this section we discuss threats to validity of our secondary literature study and applied mitigation actions. We follow the respective guidelines for managing threats to validity of secondary studies in SE by Ampatzoglou et al.~\cite{ampatzoglou2019identifying,ampatzoglou2020guidelines}. These guidelines distinguish validity threats to study selection, data, and research.

\emph{Study selection validity threats} deal with searching and including primary studies in the examined set. In the first round, inclusions in the selection stage could be reversed in the extraction stage. However, not all exclusions were verified by a second reviewer in the selection stage. This could have resulted in some papers being erroneously excluded. Still, the agreement on the papers that were checked by another reviewer (see Section~\ref{sec:selection}) indicates that this was only a moderate concern. Furthermore, because of the search verification in the second round, papers erroneously excluded could be brought back into the set. All inclusion and exclusion decisions were verified by a second extractor in the second round.

We opted for a more general search string, but it is possible that some relevant papers could use specific terms and could not be found in the initial search. Through snowballing, we would expect this threat to be reduced. We note that in the second keyword search, we found papers papers from the time period targeted in the first search. This is a potential threat to validity because it questions the repeatability of the search. We believe this may be due to updates in the databases used, either because of new publications being included, or because the search algorithms were updated. However, this limitation pertains to all literature reviews relying on database searches and is unavoidable in practice, as there is no way to verify what the databases contain nor how they execute the searches. Furthermore, we did not perform snowball sampling on the publications found in the second keyword search. This may have resulted in some publications not being found. While we acknowledge this risk, we consider it acceptable given that the keyword search and the first round of snowballing cover many relevant publications which are likely to be cited in the primary studies found after the second round of search. We already found almost as many papers with snowballing as with keyword search in the first round. This is not common for systematic literature reviews. However, the area of cognitive concepts in SE is quite fragmented. We experimented with  different search strings, but could not settle on a definitive one. Therefore, snowballing played a key role in our setting. A further limitation comes from not performing forward snowballing, which could have resulted in some papers being omitted. However, comparing the effort to the uncertain gains spoke for improving the quality of the result set in terms of its coverage of similar papers rather than an expansion of the set towards areas further away from the core topics. This way, the results of the present study should represent a more focused starting point for potential future SLR updates.

\emph{Data validity threats} deal with the extracted data set and its analysis. To reduce threats to data extraction, this activity was split among several researchers. Each extracted paper in the first round, and at least a sixth of the papers in the second round ($\geq 17\%$) was reviewed by at least one other extractor. 
Because not all papers in the second round were verified for extraction, there could be some errors in the extracted data. 
However, only minor discrepancies were found in a very small number of the papers reviewed in the second round ($\leq5\%$), indicating that such errors are few. 
Furthermore, each extraction criterion was clearly defined---i.e., using the taxonomy of cognitive concepts and assessment procedures, the SE Body of Knowledge, and simple demographic data like the year of publication. 
Overall, we had a reasonable size of primary studies (i.e., 311), and applied descriptive statistical analysis, which keeps potential threats to data analysis low.

\emph{Research validity threats} concern the overall research design. To guarantee repeatability, we documented the entire research method and make the raw data and the analysis procedures available. 
The research scope is cognition in SE. 
We do not aim to generalise our results beyond SE. 
The core contribution of this article is a taxonomy on cognitive concepts and assessment procedures to classify and discuss existing and future work. 
As this research goal requires expertise from both SE and cognitive psychology, the research was performed by a team of six SE researchers (FFA, MFE, RFE, DFU, MUN, BMA) with BLE, BNA, and JKH joining in the second phase of the literature search, and two researchers in psychology (MMA, LTE).

\section{A Taxonomy of Cognitive Concepts and Assessment Procedures in SE}
\label{sec:taxonomy}

\begin{figure}[b]
  \centering \includegraphics[width=.9\textwidth]{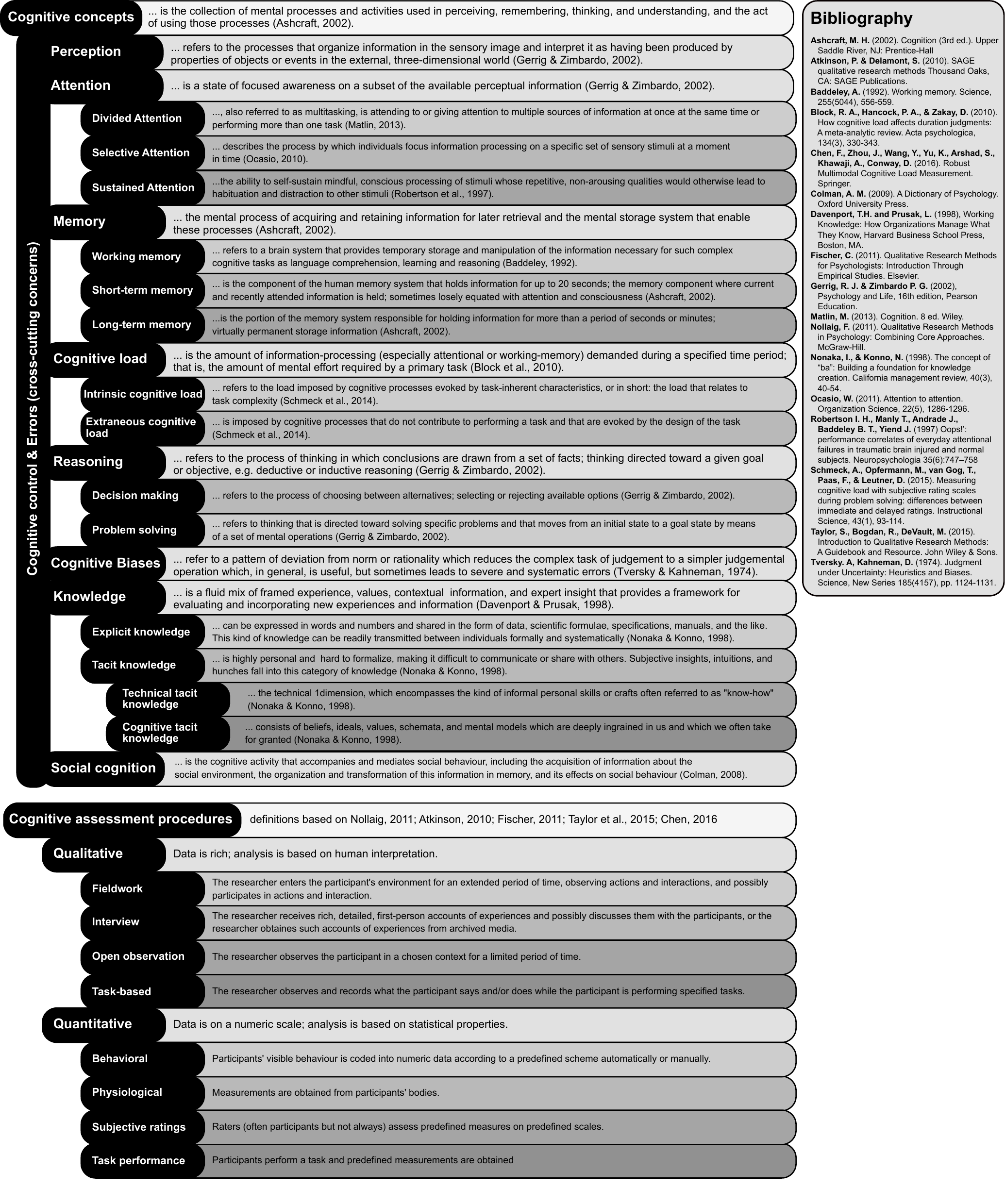}
  \caption{Taxonomy of Cognitive Research in SE. The taxonomy is divided into two main dimensions: Cognitive Concepts and Assessment Procedures.}
  \label{fig:taxonomy}
\end{figure}

To address RQ1, we present a taxonomy of cognitive concepts and assessment procedures, shown in Figure~\ref{fig:taxonomy}. The taxonomy includes two dimensions, \textit{Cognitive Concepts} and \textit{Assessment Procedures}. Cognitive Concepts define \emph{what} can be assessed, while Assessment Procedures define \emph{how} those concepts can be studied. The concepts are arranged hierarchically. The different levels of the hierarchy can be used to classify papers that differ in conceptual preciseness. Papers that study a cognitive concept in more general terms can be classified using a higher-level concept, while papers that are explicitly studying a more narrow concept can be placed in a leaf category. The hierarchically-organised concepts proceed from more basic processes on the top to higher-order ones on the bottom. Some concepts are cross-cutting, spanning the range of all other concepts. Each combination of a cross-cutting concern with another concept constitutes a particular kind of variation on a cognitive process.

The taxonomy serves as an analytical tool to characterise research on cognition in SE and shows the extent to which articles focus on specific topics. We argue that it can inform practitioners by giving an overview of the field and helping to judge how aspects of cognition can affect them and their organisations. Furthermore, it helps to identify knowledge gaps and point researchers to new investigations.

\subsection{Cognitive Concepts}

The cognitive concepts in the taxonomy capture a collection of important processes and activities of individual cognition at different levels, providing an answer to the classification part of RQ1. We briefly expand on each concept below.

\subsubsection{Perception}

The processes that organise and interpret sensory information in order to understand it or the environment it presents (e.g., visual information regarding a three-dimensional world) are called \emph{perception}. Through perception, sensory stimuli are interpreted as having been produced by properties of objects or events in the environment. Perceptual processes are not only passive reception of sensory information, but are shaped by other cognitive processes, such as memory and attention. Information processed by perception includes sensory information---vision, sound, touch, taste, and smell---but also social information, such as understanding of several aspects of speech and faces, as well as perception of time, agency, and familiarity.

\subsubsection{Attention}

Senses (e.g., vision, hearing, and touch) are systems involved in sensation, the physical process where stimuli (e.g., light, sound, and pressure) are collected and provided for perception. \emph{Attention} is a state of focused awareness on a subset of the available perceptual information. In other words, attention refers to the selection of what perceptual information an individual primarily focuses on.

This concept is divided into three sub-concepts: \emph{selective}, \emph{divided}, and \emph{sustained} attention. \emph{Selective attention} is the process of directing the limited resources of attention towards specific stimuli while ignoring other stimuli. This enables humans to focus on what is deemed important in the situation or task. \emph{Divided attention} refers to the ability to integrate multiple stimuli in parallel. It enables paying attention to multiple things at the same time and is required for many everyday tasks as well as complex tasks in the workplace. For example, a software developer may have to juggle between information coming from several different programs to track down a program fault. \emph{Sustained attention} refers to the ability to focus on a specific task or thought for an extended period of time, while avoiding being distracted by other stimuli. This is especially important in complex, knowledge-intensive tasks that require concentration over long periods of time. It is also involved in learning-related processes, such as program comprehension.

\subsubsection{Memory}

The mental processes of acquiring and retaining information for later retrieval, as well as the mental storage system that enables these processes, are referred to as \emph{memory}. Memory can be divided into sub-concepts based on its role and how long information is retained. \emph{Working memory} is a brain function providing temporary storage and manipulation of information needed for complex tasks such as language comprehension, learning, and reasoning. \emph{Short-term memory} holds current and recently attended information briefly, in the range of seconds or minutes. \emph{Long-term memory} holds information for longer periods, being virtually permanent.

\subsubsection{Cognitive Load}

The amount of information-processing required, especially in terms of attentional and working memory, differ between tasks. The information-processing needs during a specific time period is referred to as \emph{cognitive load}, and can be divided into the load that relates to task complexity (\emph{intrinsic cognitive load}) and the load that relates to cognitive processes that do not contribute to performing the task itself (\emph{extraneous cognitive load}).

\subsubsection{Reasoning}

Thinking that is directed toward a given goal or objective is referred to as \emph{Reasoning}. This includes \emph{problem-solving}, which is thinking directed toward solving specific problems, moving from an initial state to a goal state by means of mental operations. Another kind of reasoning is \emph{decision-making}, which refers to the process of choosing between alternatives, selecting or rejecting available options, and potentially creating new options.

\subsubsection{Cognitive Biases}

\emph{Cognitive biases} refer to patterns of deviation from norm or rationality which originate from reduction of complex judgement tasks to simpler judgement operations. In general, such simplifications are useful and serve to produce quick results that often are accurate enough but that sometimes lead to severe systematic errors. Cognitive biases can be seen as particular kinds of errors (see below) and originate from the limitations of other cognitive processes.

\subsubsection{Knowledge}

Experience, values, contextual information, and expert insight form a framework referred to as \emph{knowledge}. This framework is used to evaluate new experiences and information, but the framework itself is also altered in the process. Specific kinds of knowledge are identified. \emph{Explicit knowledge} can be expressed as words and numbers, and is readily shared and transmitted between individuals. In contrast, \emph{tacit knowledge} is personal, hard to formalise, and difficult to communicate or share with others. Such knowledge can be divided into \emph{technical tacit knowledge}, or ``know-how,'' and \emph{cognitive tacit knowledge}---i.e., those beliefs, ideals, values, schemata, and mental models which are deeply ingrained in individuals and often taken for granted.

\subsubsection{Social Cognition}

Perceptual processes provide information about the social context of an individual, such as facial expressions, actions, and other overt behaviour of other individuals. The individual also displays overt behaviours to others in the social context. The cognitive activity that accompanies and mediates social behaviour, including acquisition of information about the social environment, organisation and transformation of this information in memory, and its effects on the individual's behaviour are referred to as \emph{social cognition}. It is a cross-cutting concept: it includes many levels of cognition and interacts with different concepts, from lower levels of the taxonomy to higher-order functions.

\subsubsection{Cognitive Control}

\emph{Cognitive control}, also known as \emph{executive function} is a set of cognitive processes that work to  govern behaviour. This includes both basic cognitive processes, such as attentional control, cognitive inhibition, working memory, and cognitive flexibility, and higher-order functions such as planning, reasoning, and problem-solving. Cognitive control is not a single process but multiple processes that help monitor and select behaviours that facilitate attainment of goals. It is thus a cross-cutting concept in the taxonomy.

\subsubsection{Errors}

Cognitive processes, even when functioning correctly, can produce irrational results. Such \emph{errors} originate from, for example, perceptual distortions, inaccurate judgement, or other irrationalities.

\subsection{Assessment Procedures}

\emph{Assessment procedures} are different means of obtaining systematic information about the quality of each cognitive concept in a particular context. This ranges from measures obtained from measurement devices to observations made by the researcher. We divide the research methods into \emph{quantitative} and \emph{qualitative}, and base their characterisation on the \textit{Bibliography} reported in Figure~\ref{fig:taxonomy}. This provides an answer to the assessment part of RQ1.

\subsubsection{Qualitative}

When the assessment procedure is based on rich data and human interpretation, we refer to it as \emph{qualitative}. In such research, the researcher interprets information without necessarily decomposing it into its constituent parts or assigning numbers to the observations. Qualitative assessment procedures include \emph{fieldwork}, where the researcher enters the environment where the assessment is to be made, possibly participating in actions and interaction, for an extended period of time; \emph{interviews}, where the researcher obtains first-person or documented accounts of experiences either from an individual or as a group (e.g., focus group); \emph{task-based} procedures, where the researcher observes and records the conduct of specified tasks; and \emph{open observation}, where assessment is based on observation in a chosen context for a limited time.

\subsubsection{Quantitative}

When the assessment procedure is based on data on a numeric scale and analysis of statistical properties, we refer to it as \emph{quantitative}. Such research includes \emph{task performance} assessments, where participants perform a task and predefined measurements are obtained; \emph{physiological}, where measurements are obtained from participants' bodies; \emph{subjective ratings}, where raters, who may or may not be the participants themselves, rate artefacts using predefined scales; and \emph{behavioural}, where participants' overt behaviour is coded into numeric data according to a predefined scheme, either automatically or manually by specially trained raters.

\section{Literature Survey Results}
\label{sec:results}

Research on cognition in SE has seen attention for more than five decades. This body of literature covers several research approaches, SE knowledge areas, and study contexts, all of which are detailed in this section. We use our taxonomy (see Figure~\ref{fig:taxonomy}), together with the SE Body of Knowledge (SWEBOK)~\cite{swebok}, to illustrate the foci of the last five decades of research on cognitive concepts in SE and provide answers to RQ2 and its sub-questions. At the same time, these results act as a validation of the taxonomy of cognitive concepts and assessment procedures. We show how the taxonomy is able to organise and situate the research in terms of concepts from and related to cognitive psychology. 
In the figures below, we omit the concepts \emph{Cognitive control} and \emph{Social cognition}, as the former includes only one publication from 1988, and the latter includes only seven publications between 2008 and 2020. However, we analyse these concepts and the related publications where applicable. We also omit the super-category \emph{Errors} when the results only contain articles in the sub-category \emph{Cognitive biases}.

\subsection{Research Evolution, Context, and Type}

\begin{figure}[b]
\centering
  \includegraphics[width=0.8\linewidth]{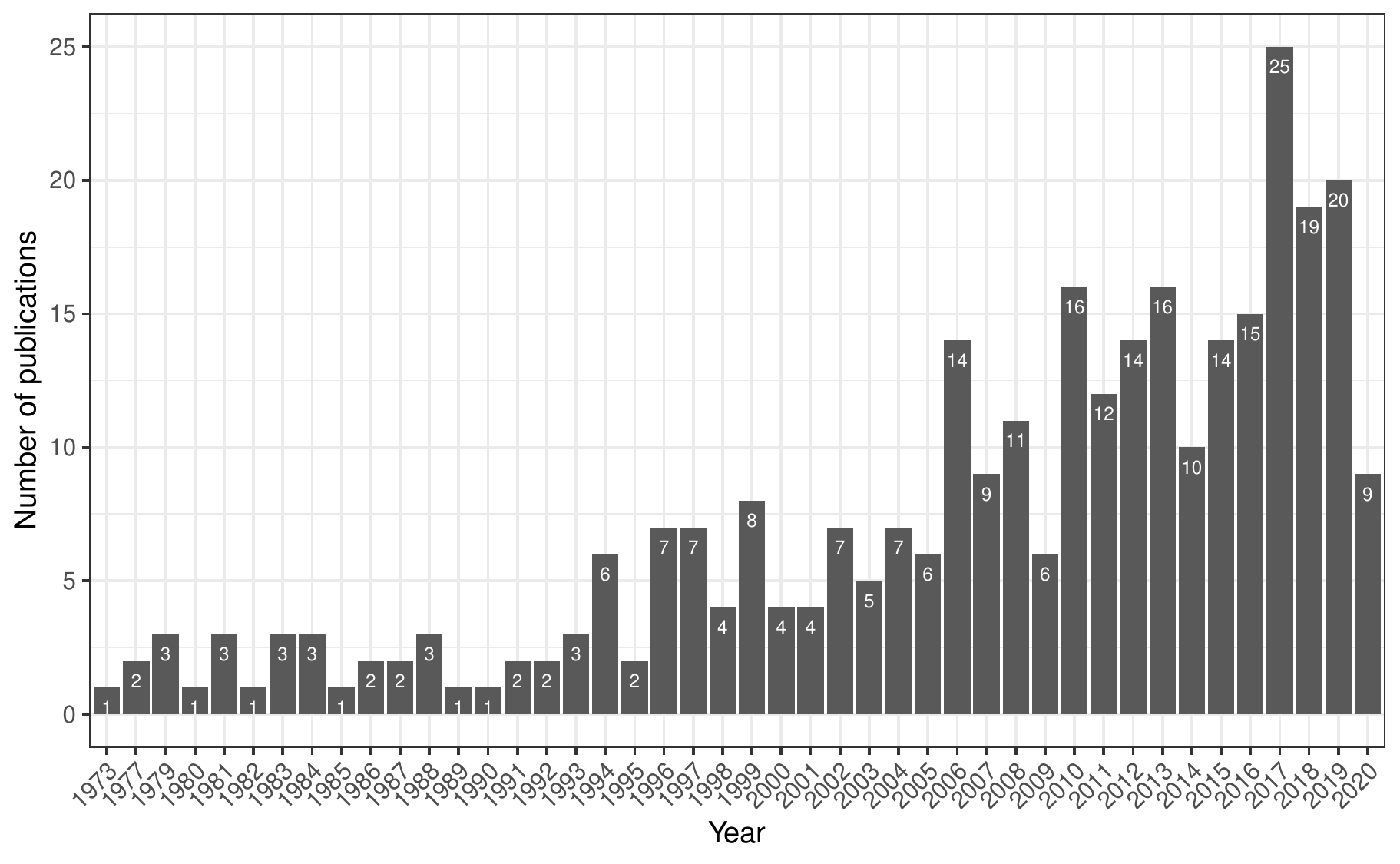}
  \caption{Distribution of publications per year. The data ranges from 1973 to May 2020.}\label{fig:yearly_distribution}
\end{figure}

We first examine the contexts in which research on cognitive concepts in SE has been performed (RQ2.1) and the evolution of that research over time (RQ2.2). Figure~\ref{fig:yearly_distribution} shows the number of publications per year. The earliest paper found was from 1973, reporting on a psychological evaluation of two conditional constructs commonly found in many programming languages, if-goto and if-then-else~\cite{sime1973}. The paper belongs to the long-standing family of studies in program comprehension. Interestingly, the authors are affiliated with a department of psychology, showing how that field has had an influence on the body of knowledge related to SE for decades. Nevertheless, as noted earlier, psychology cannot be said to have become part of the foundations of SE research.

\begin{figure}
\centering
\begin{subfigure}{0.45\textwidth}
\centering
  \includegraphics[width=.8\textwidth]{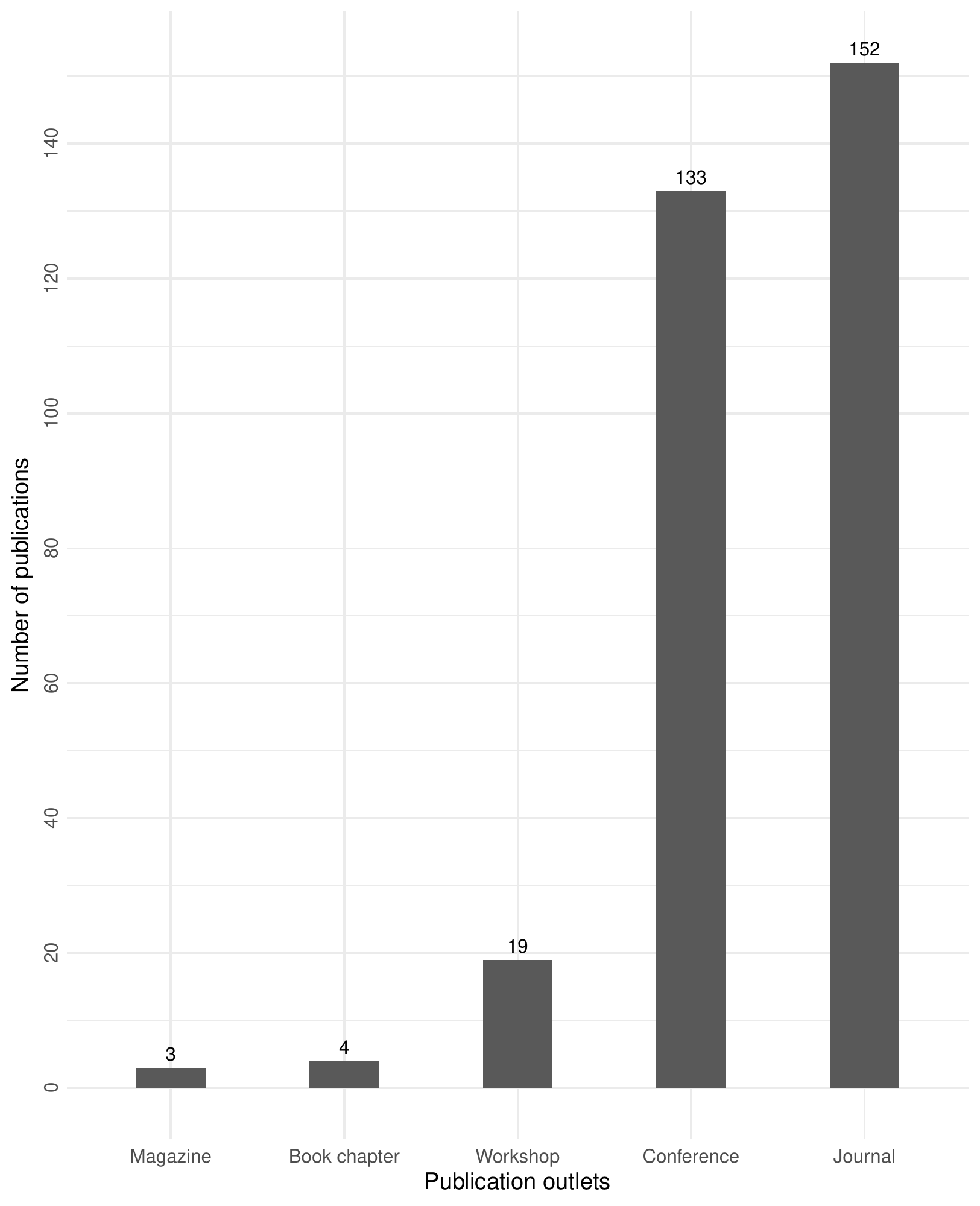}
  \caption{Publications per scientific outlets.}
  \label{fig:outlets}
\end{subfigure}
\begin{subfigure}{0.45\textwidth}
\centering
  \includegraphics[width=.8\textwidth]{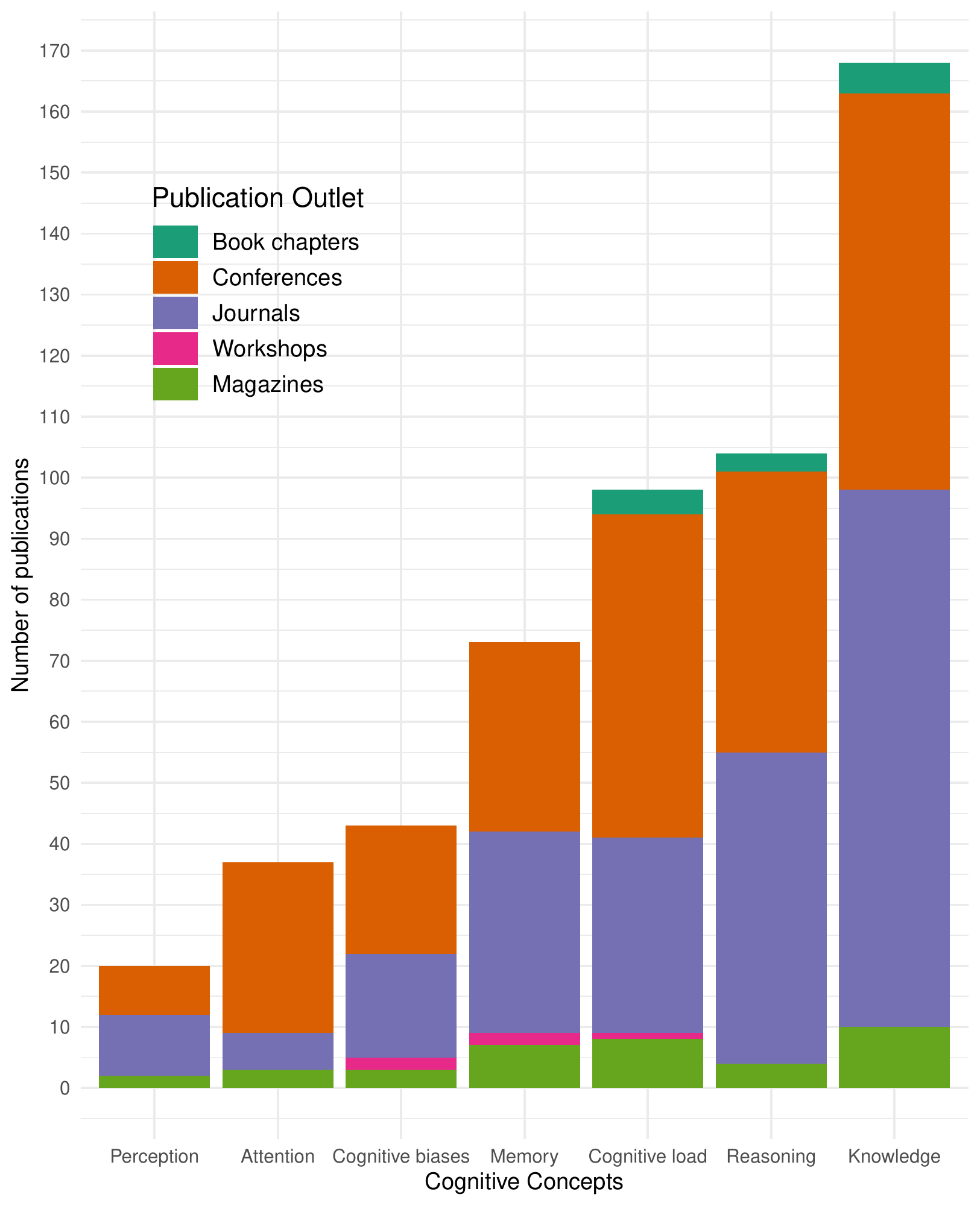}
  \caption{Publications per outlets and cognitive concepts. }
  \label{fig:outlets_concepts}
\end{subfigure}
\begin{subfigure}{.45\textwidth}
\centering
  \includegraphics[width=.8\textwidth]{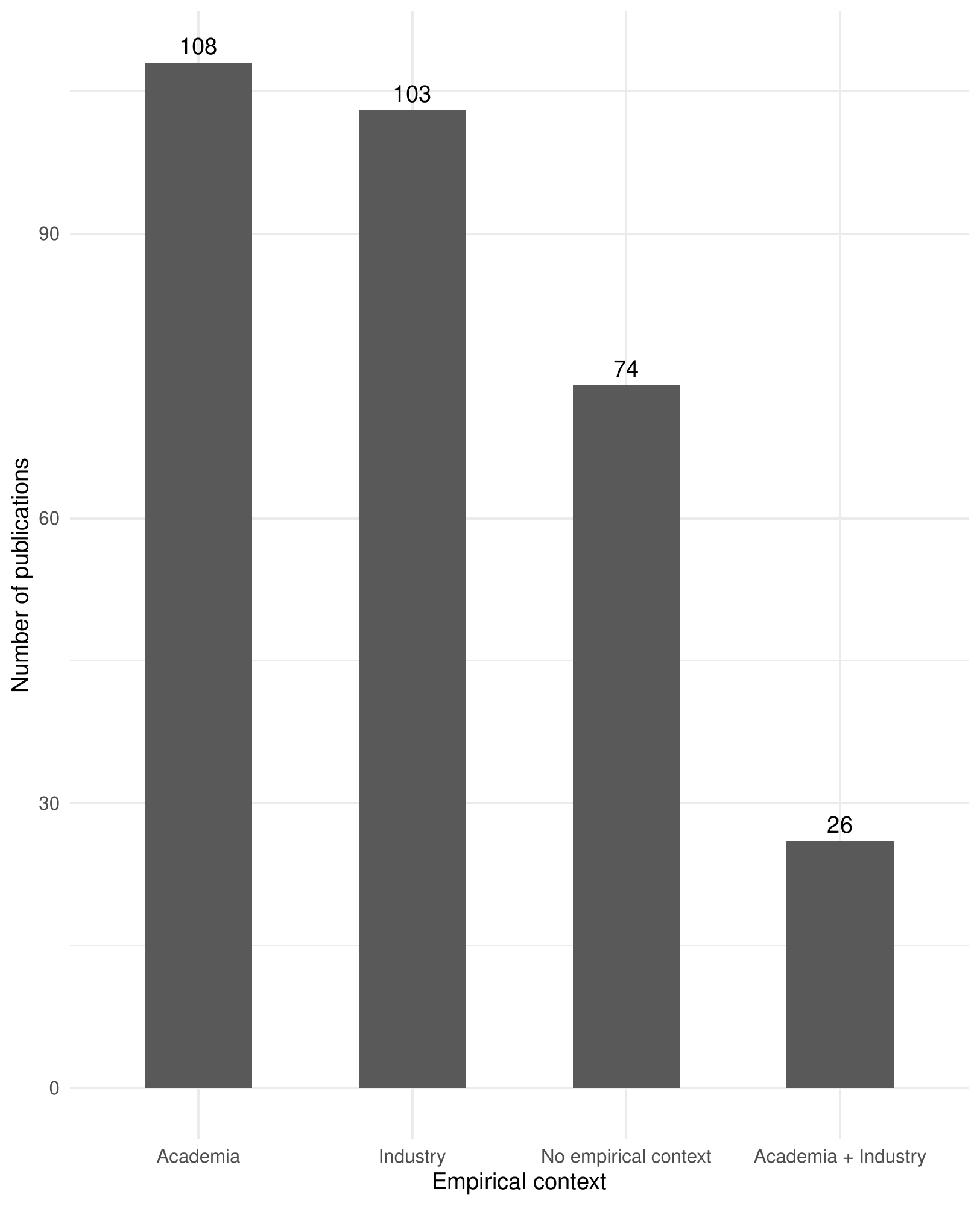}
  \caption{Publications per research context.}
  \label{fig:academia_industry_distribution}
\end{subfigure}
\begin{subfigure}{0.45\textwidth}
\centering
  \includegraphics[width=.8\textwidth]{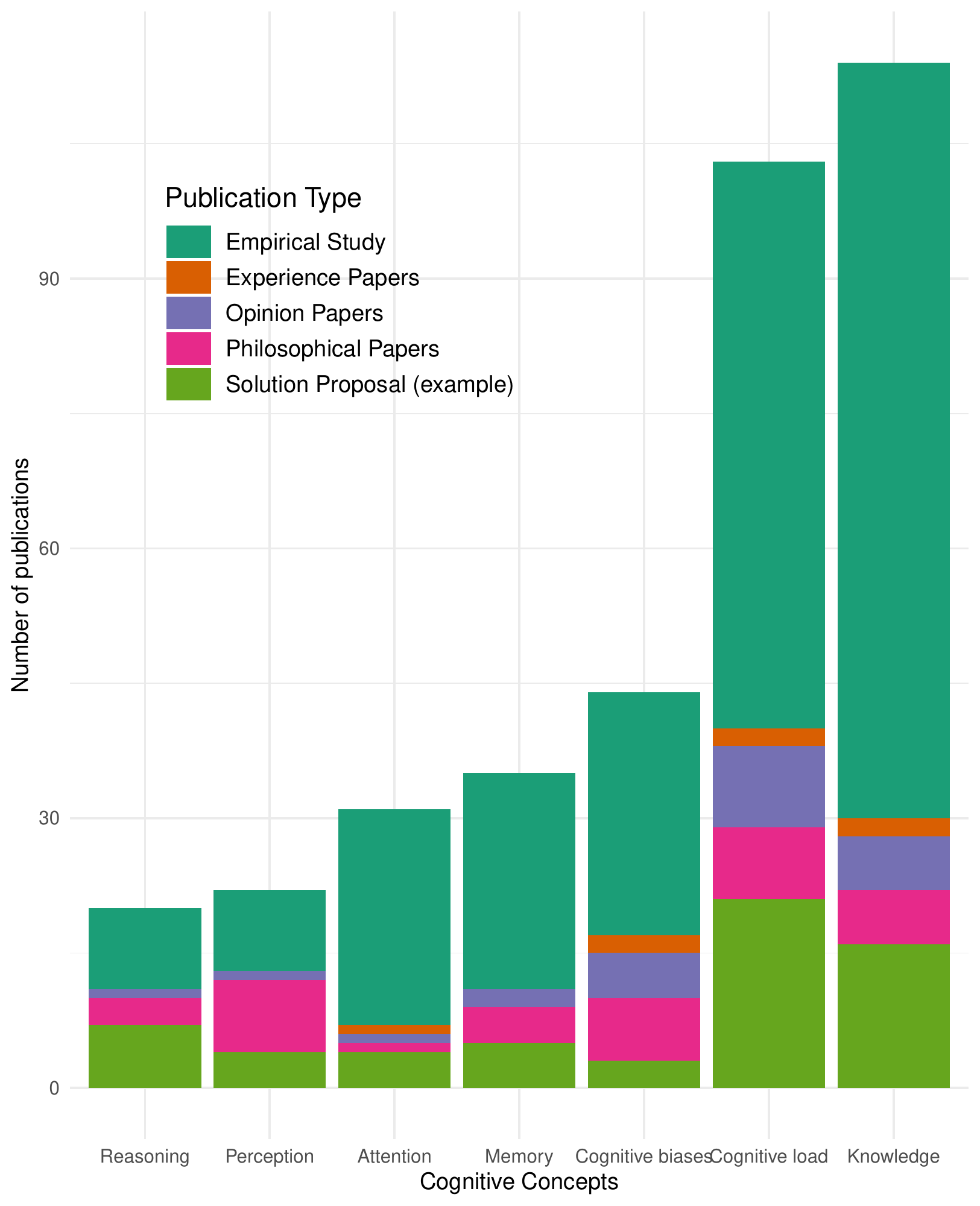}
  \caption{Publications per study types and cognitive concept.}
  \label{fig:paper_type}
\end{subfigure}

\caption{Distribution of SE publications according to outlets, paper type, and cognitive concepts.}
\label{fig:outlets_all}
\end{figure}

We analysed the outlets that publish work on cognitive concepts, the context in which published research takes place, as well as the publication types (Figure~\ref{fig:outlets_all}).
In particular, Figure~\ref{fig:outlets} shows the number of publications in different types of scientific outlets. The publications can be mainly found in journals and conferences, as expected in this field. Workshop and magazine articles, together with book chapters, make up roughly 10\% of the total set of primary studies.
Different cognitive concepts are covered to a different extent in different types of outlets (see Figure~\ref{fig:outlets_concepts}). \emph{Cognitive load}, \emph{Memory} and \emph{Cognitive biases} are each covered in workshop papers, while the other areas are not discussed in workshop papers. Papers regarding all the cognitive concepts, with the exception of \emph{Attention} are published in similar numbers in journals and conference proceedings. 

Figure~\ref{fig:academia_industry_distribution} shows the number of publications according to the context in which the studies were conducted. Studies performed in an academic context, which typically means having student participants, are only slightly more common than studies in an industry context. Studies in which both an academic and industry context are utilised are the least common. There are also studies with no empirical context---these are theoretical, philosophical, or position papers that do not include an empirical component but can contribute to conceptual and theoretical development in the area.

Figure~\ref{fig:paper_type} shows which kind of papers, according to the Wieringa et al.\ SE publications taxonomy~\citep{wieringa_requirements_2005}, cover the different cognitive concepts. The \emph{Knowledge} concept is most addressed, through a large number of empirical studies. However, \emph{Cognitive load}, the second-most addressed concept overall, had more solution proposal, philosophical, and opinion papers. The third-most addressed concept was \emph{Cognitive biases}, while \emph{Memory}, \emph{Attention}, \emph{Perception}, and \emph{Reasoning} were the least addressed, in that order. A large majority of all papers were empirical studies. Solution proposals and philosophical papers were frequent but much less so than the empirical papers. Some opinion and experience papers were found; the former concerning all concepts, and the latter concerning \emph{Knowledge}, \emph{Cognitive load}, \emph{Cognitive biases}, and \emph{Attention}. Overall, this indicates that the research represented by the papers in our result set is grounded in empirical evidence, that there is some philosophical grounding in all the concepts, and that some solution proposals have been put forward.

\begin{figure}[b]
  \centering
  \includegraphics[scale=.57]{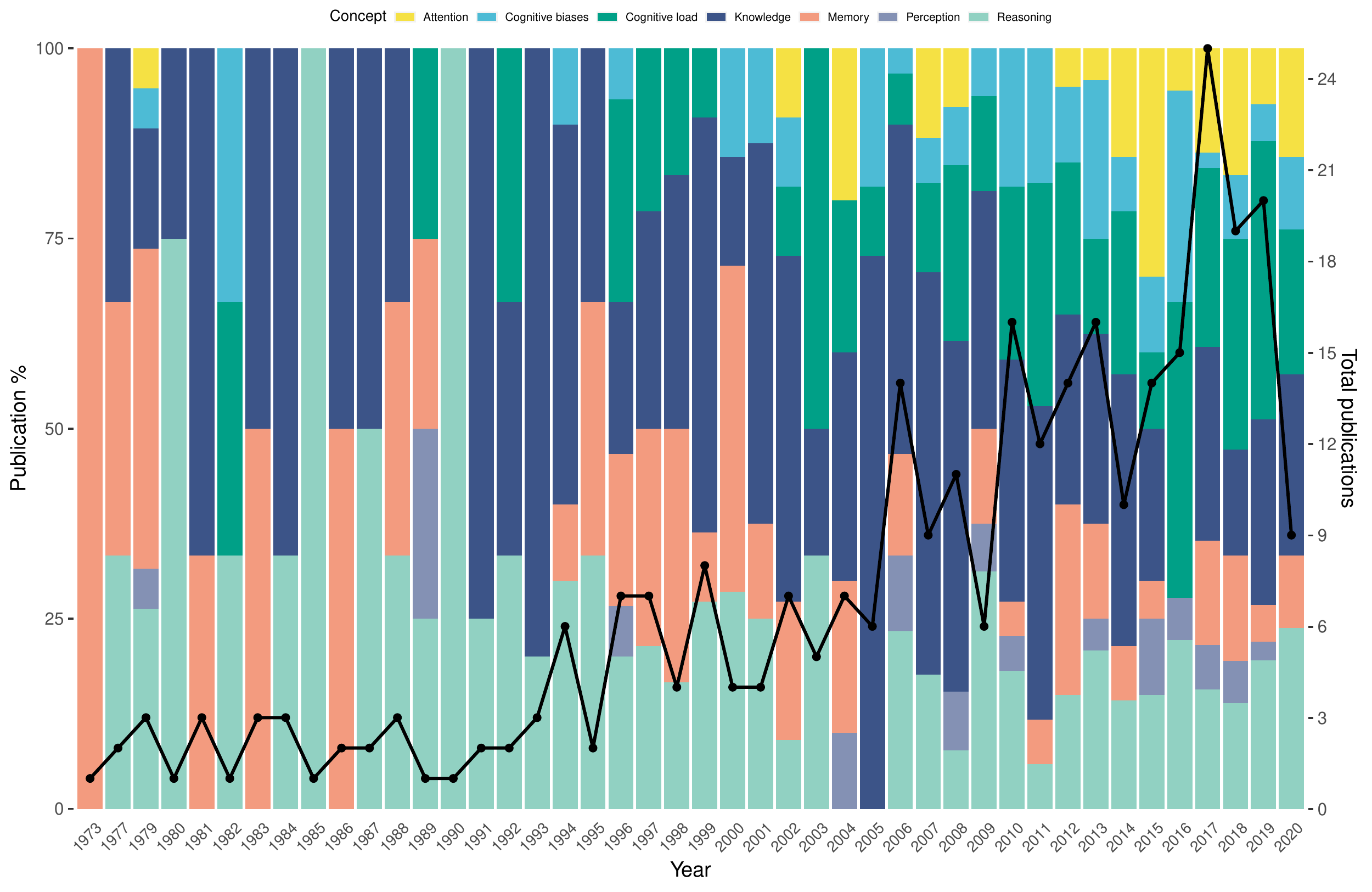}
  \caption{Percentage of publications covering the different cognitive concepts for each year. The black line indicates the total number of publications. No publications for years 1974--1976 and 1978. A publication can cover more than one concept.}
  \label{fig:taxonomy_trends}
\end{figure}

Research on different cognitive concepts has evolved over time, as shown in Figure~\ref{fig:taxonomy_trends}. Studies related to \emph{Reasoning}, \emph{Perception}, \emph{Memory}, and \emph{Knowledge} are published regularly since 1973, and are prevalent until the early 1990s while other areas have longer gaps between years of publication during the period we consider. Studies dealing with \emph{Cognitive load} and \emph{Cognitive biases} first appeared in the early 1980s but were in focus after the year 2000. In 2016, more than 75\% of papers were published in these areas.
Research on \emph{Attention}, after being introduced in the literature in 1979, and with some occurrences in the first part of the 2000s, gained traction in 2012 and in 2015 almost one third of the published primary studies dealt with this concept. \emph{Social cognition} (omitted from the figure) has seen recent attention, but it is too early to say if this will lead to a longer-term research focus. \emph{Cognitive control} (also omitted from the figure) appears not to have gained such focus, with only one publication from 1988.

\subsection{Coverage of Concepts and SWEBOK Areas}
\begin{figure}
  \centering
  \includegraphics[width=.8\textwidth]{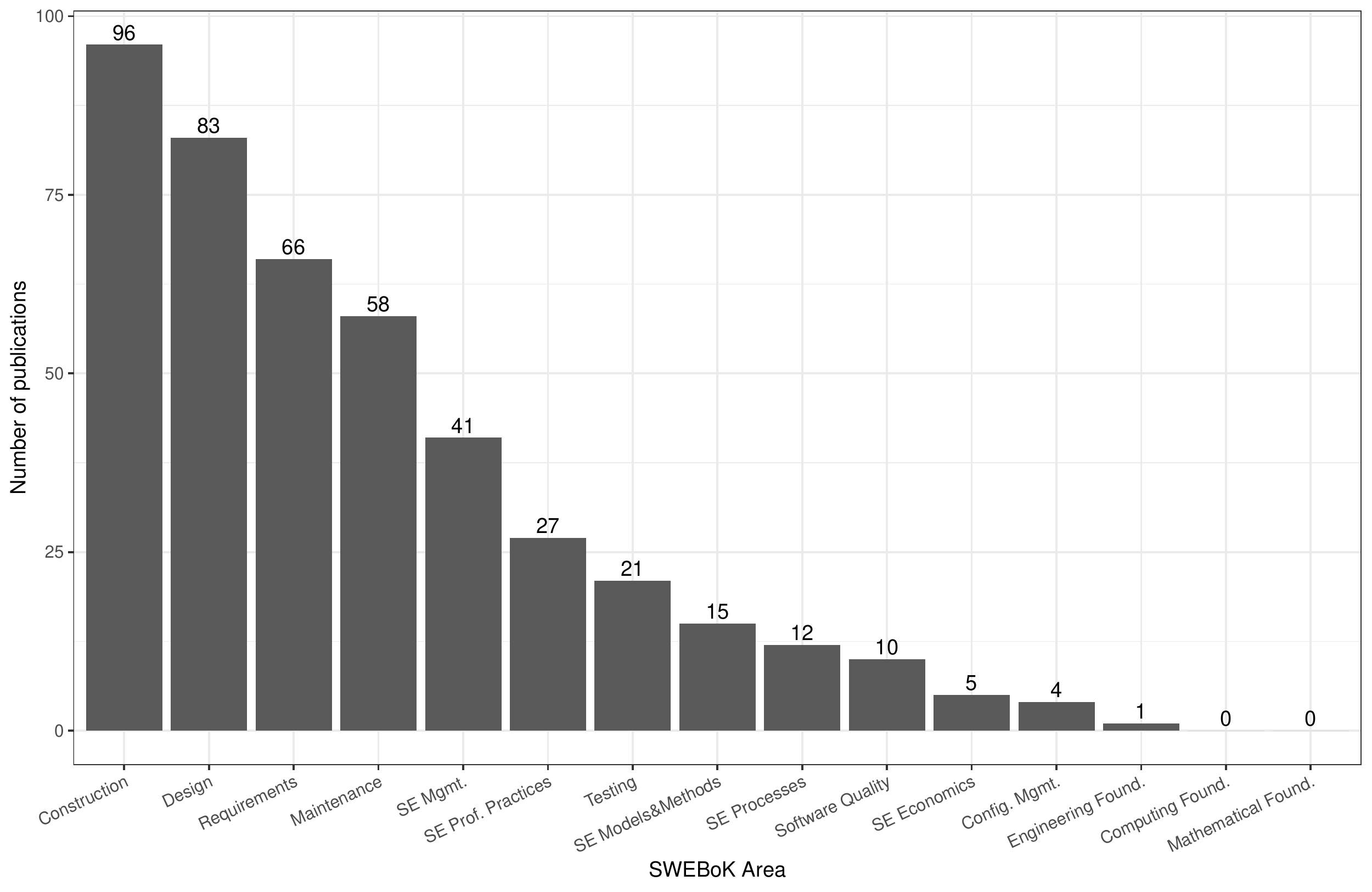}
  \caption{Number of publications according to SWEBOK areas.}
  \label{fig:swebok_distribution}
\end{figure}

We now turn to examining the SE knowledge areas that have been studied under the lens of cognitive concepts, addressing RQ2.3. Figure~\ref{fig:swebok_distribution} shows the distribution of publications according to SWEBOK areas. The largest number of publications are found in Software Construction, Design, Requirements, and Maintenance, each with at least 50 publications\footnote{A paper can cover one or more SWEBOK area.}. There are relatively few papers on Software Testing, Process, and Quality.
There is no publication in the Computing and Mathematical foundation, and a single publication in the Engineering foundation\footnote{Due to the low number of publications, we do not report results about these areas in the rest of this paper.}.

\begin{figure}[b]
\begin{subfigure}{.32\textwidth}
  \centering
  \includegraphics[width=\linewidth]{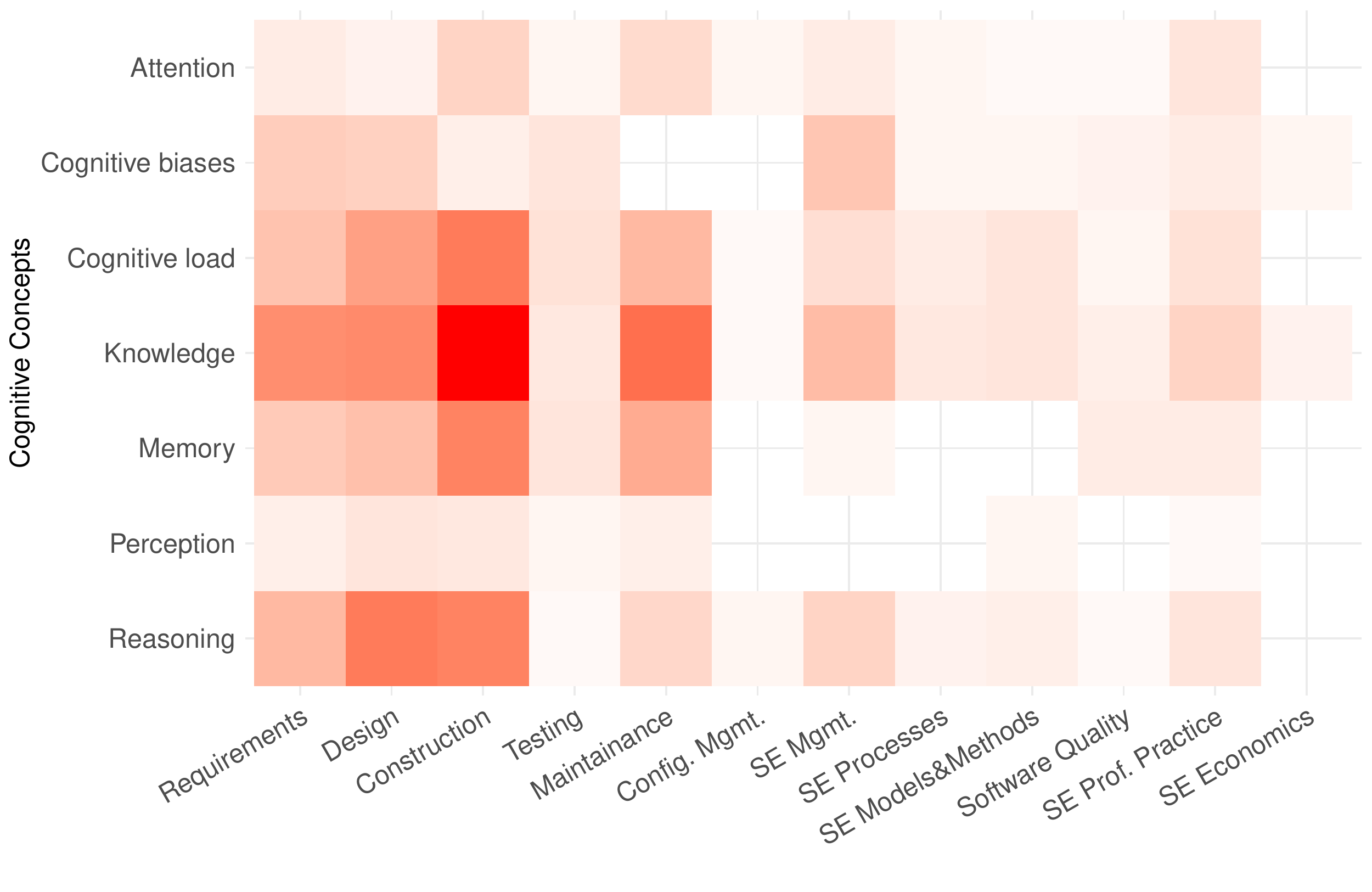}
  \caption{All publications.}
    \label{fig:taxonomy_swebok_all}
\end{subfigure}%
\begin{subfigure}{.32\textwidth}
  \centering
  \includegraphics[width=\linewidth]{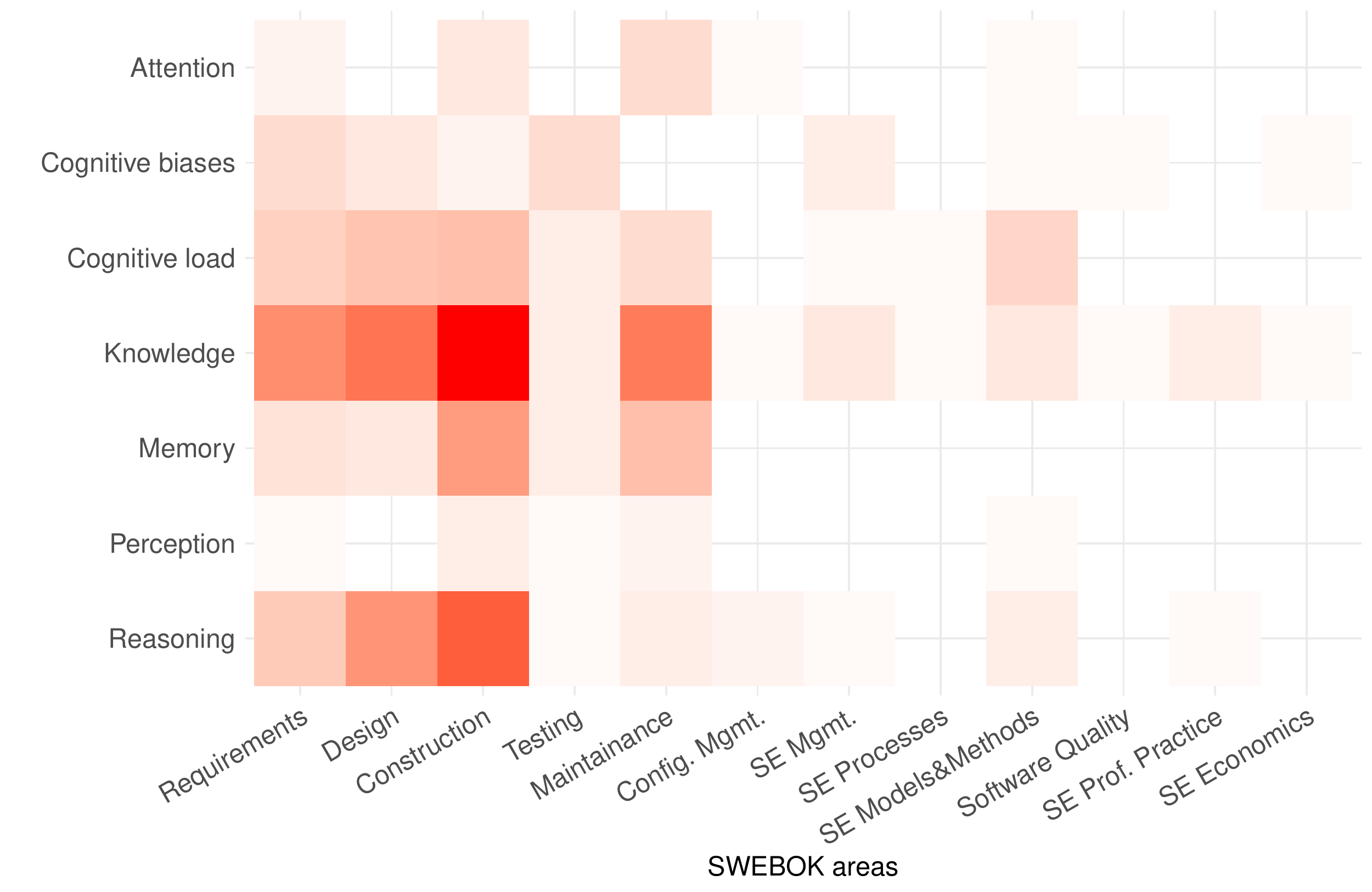}
  \caption{Publications in Academic context.}
  \label{fig:taxonomy_swebok_academia}
\end{subfigure}
\begin{subfigure}{.32\textwidth}
  \centering
  \includegraphics[width=\linewidth]{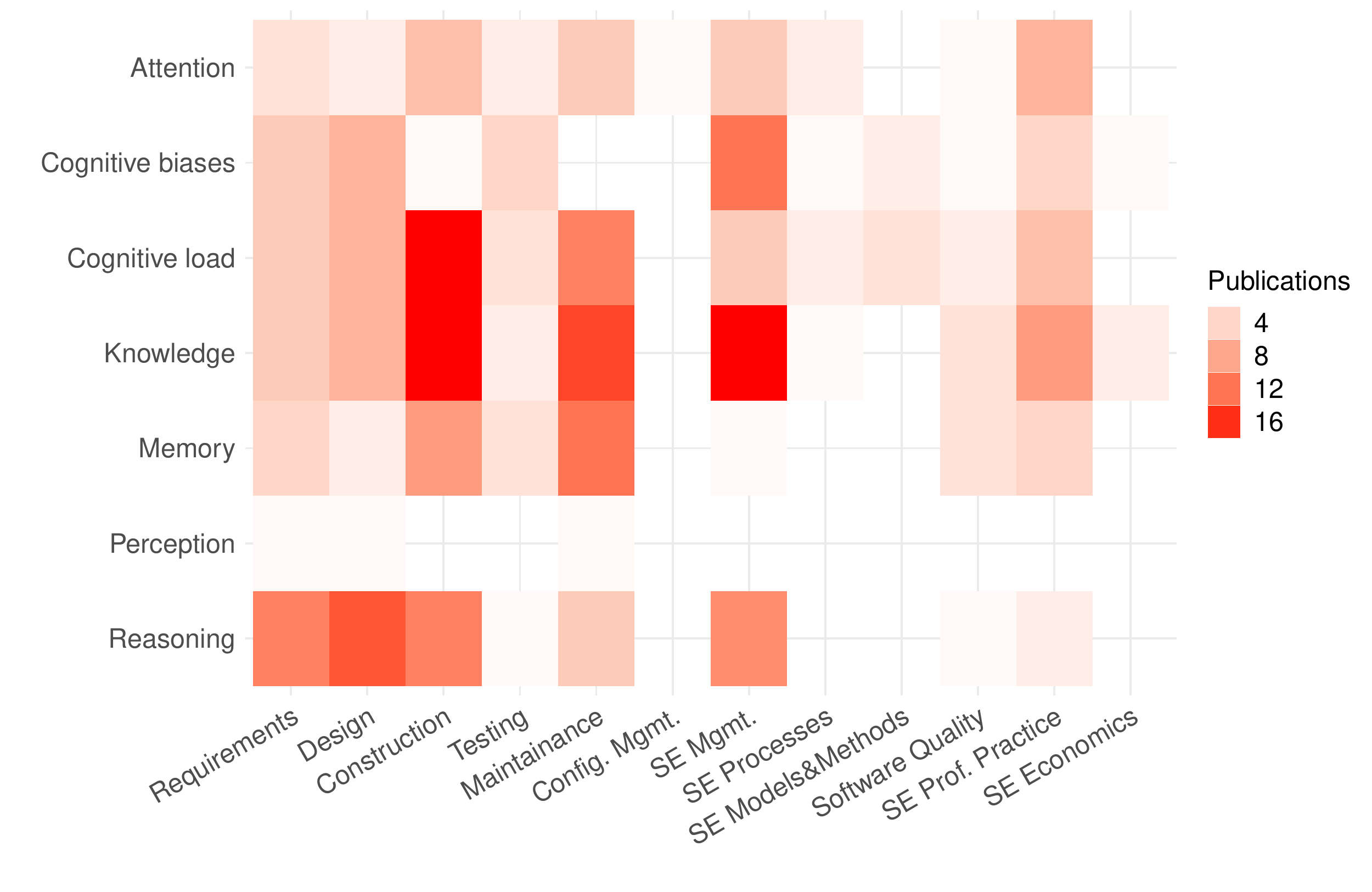}
  \caption{Publications in Industrial context.}
  \label{fig:taxonomy_swebok_industry}
\end{subfigure}
\caption{Publications characterised by Cognitive Concepts in the different SWEBOK areas. Computing, Mathematical, and Engineering foundations are excluded as there is only one study published in these areas.}
 \label{fig:taxonomy_swebok}
 \end{figure}

The concept coverage differs between SWEBOK areas (see Figure~\ref{fig:taxonomy_swebok}). Four central SE life-cycle areas---Requirements, Design, Construction, and Maintenance---are most often covered by the \emph{Knowledge} concept and, to a lesser extent, by \emph{Cognitive load}, \emph{Reasoning}, \emph{Memory}, and \emph{Cognitive biases}. \emph{Knowledge} is a broad concept and it is inevitable that it is covered in many primary studies. No concept is predominant in the area of Software Testing, for which \textit{Reasoning}, \textit{Perception}, and \textit{Attention} are scarcely covered.

\begin{figure}
  \centering
  \includegraphics[width=.8\textwidth]{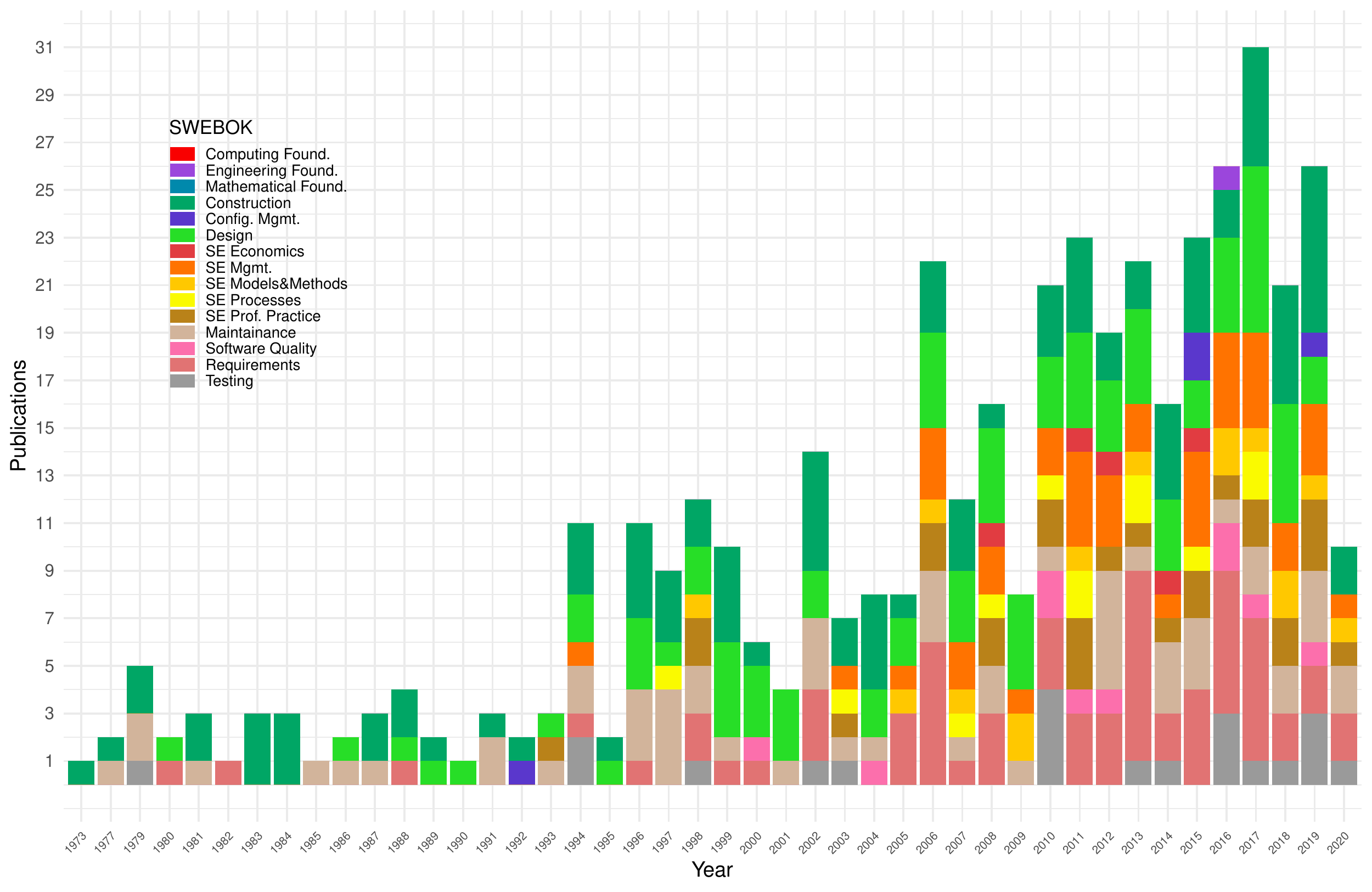}
  \caption{Distribution of publications covering the SWEBOK areas per year.}
  \label{fig:years_swebok}
\end{figure}

Part of this result can be attributed to the fact that there are, overall, more papers in the main life-cycle areas in SWEBOK than in areas such as SE Management or Economics. However, this does not fully explain the lack of research in Testing, Configuration Management, Processes, or Quality---all areas with large amounts of publications. It is also worth noting that SE Professional Practice, which explicitly covers human aspects such as individual cognition and problem-solving, has very few papers covering any cognitive concept at all. 
The lack of research in these areas can be partly explained by papers not being classified appropriately when their main focus is on a life-cycle area.

Examining the areas and concepts more closely, we see how the taxonomy organises the research. \emph{Attention} stands out as being investigated mostly in Software Construction, Maintenance, and to a lesser extent in Professional Practice. This shows that the taxonomy is able to capture an important cognitive process that is relevant to performing technical, and often programming-related tasks. Also with \emph{Reasoning}, we see a difference in how the cognitive concept is emphasised in different SE areas. While research on \emph{Attention} focuses on construction and maintenance tasks, \emph{Reasoning} covers more of the earlier life-cycles stages such as Requirements and Design. This is in line with the definition of the concepts, since \emph{Reasoning} includes problem-solving and decision-making activities that are relevant for these SWEBOK areas. It may be more difficult, and therefore less common, to investigate \emph{Attention} in such areas.

Figure~\ref{fig:years_swebok} shows the distribution of publications per year according to the SWEBOK areas. The published work in Software Testing has increased slightly since 2010, indicating a possibility that cognition will be better covered in this area in the future compared to our findings. Research in Software Quality has also grown recently, but the number of publications in this area is still small and papers are sporadically published.

The research on cognitive concepts has had focal changes over the observed time period. We observed changes in different parts of the taxonomy. Focus on \emph{Attention} has increased over the observation period. However, \emph{Selective}, \emph{Divided} and \emph{Sustained Attention} are not commonly reported. For the latter, all research is quite recent. We believe this indicates a lack of conceptual clarity in research investigating \emph{Attention}.
The role of attention in several software development tasks is important.
Therefore, research in this area needs to be more grounded in established conceptualisations and specify more precisely what aspect of attention is investigated.

\emph{Memory} has been researched more during 1973-1988 and after that the work in this area seems to have declined. \emph{Working memory} has sporadic coverage over the whole area and more regular coverage in the last ten years, but there are still few primary studies on this concept. \emph{Short-term memory} was covered more in the 1970s and 1980s, and again at the end of the 1990s and start of 2000s, but there have been no papers published on this concept in the last ten years. \emph{Long-term memory} is covered in some papers, with a concentration at the end of the 1990s and start of 2000s, but seems to have declined since then.

\emph{Cognitive load} has become more regularly addressed and the relative amount of research has grown. Cognitive load is a concept that feels intuitive to address, as practitioners and researchers alike can relate to the thought of being overloaded during a demanding task. However, we see room for improvement in the use of this concept. First, it is most commonly researched as a general concept without much regard to the underlying explanation of how cognitive load arises. This can be observed in the very sparse occurrences of research that explicitly examines the \emph{Intrinsic} and \emph{Extrinsic cognitive load} concepts. Thus, studies do not seem to conceptually differentiate whether the load is related to task complexity or evoked by the task design. Second, as cognitive load is related especially to working memory, we would expect to see many papers on cognitive load to also address \emph{Memory}. However, that concept is in decline, and only 20\% of cognitive load papers consider memory. Studies could gain explanatory power by considering the relationship between cognitive load and memory, and by building on related studies in cognitive psychology.

\emph{Perception} is regularly addressed during the observation period. There are not many papers on this concept but the regularity of publications has increased to some extent. It can be useful to consider the human perceptual processes and the concepts around them more specifically in SE research.

Research on \emph{Reasoning} covers the whole period, but only sporadically. The \emph{Problem-solving} concept is frequently addressed and there is a relatively large number of publications, which seems to be declining. \emph{Decision-making} is more regularly addressed in recent years, but does not represent a particularly large portion of the studies.

The \emph{Cognitive biases} (\emph{Errors}) concept has been addressed more regularly since 2005. However, it seems too early to say whether it will grow to cover a significant portion of the research on cognition in SE.

\emph{Knowledge} is regularly and frequently addressed over the whole period. Similarly to \emph{Attention}, the more precise division into sub-concepts is less common. \emph{Explicit knowledge} is sporadically addressed and seems to have decreased in recent years. \emph{Tacit knowledge} is covered in some papers since 1998, but sporadically and in small numbers. \emph{Technical tacit knowledge} has been covered sporadically throughout the period but has become more sparse between 2010 and 2020. \emph{Cognitive tacit knowledge} has been addressed very sparsely.

\subsection{Coverage of Assessment Procedures}
\begin{figure}
\begin{subfigure}{.32\textwidth}
  \centering
  \includegraphics[width=\linewidth]{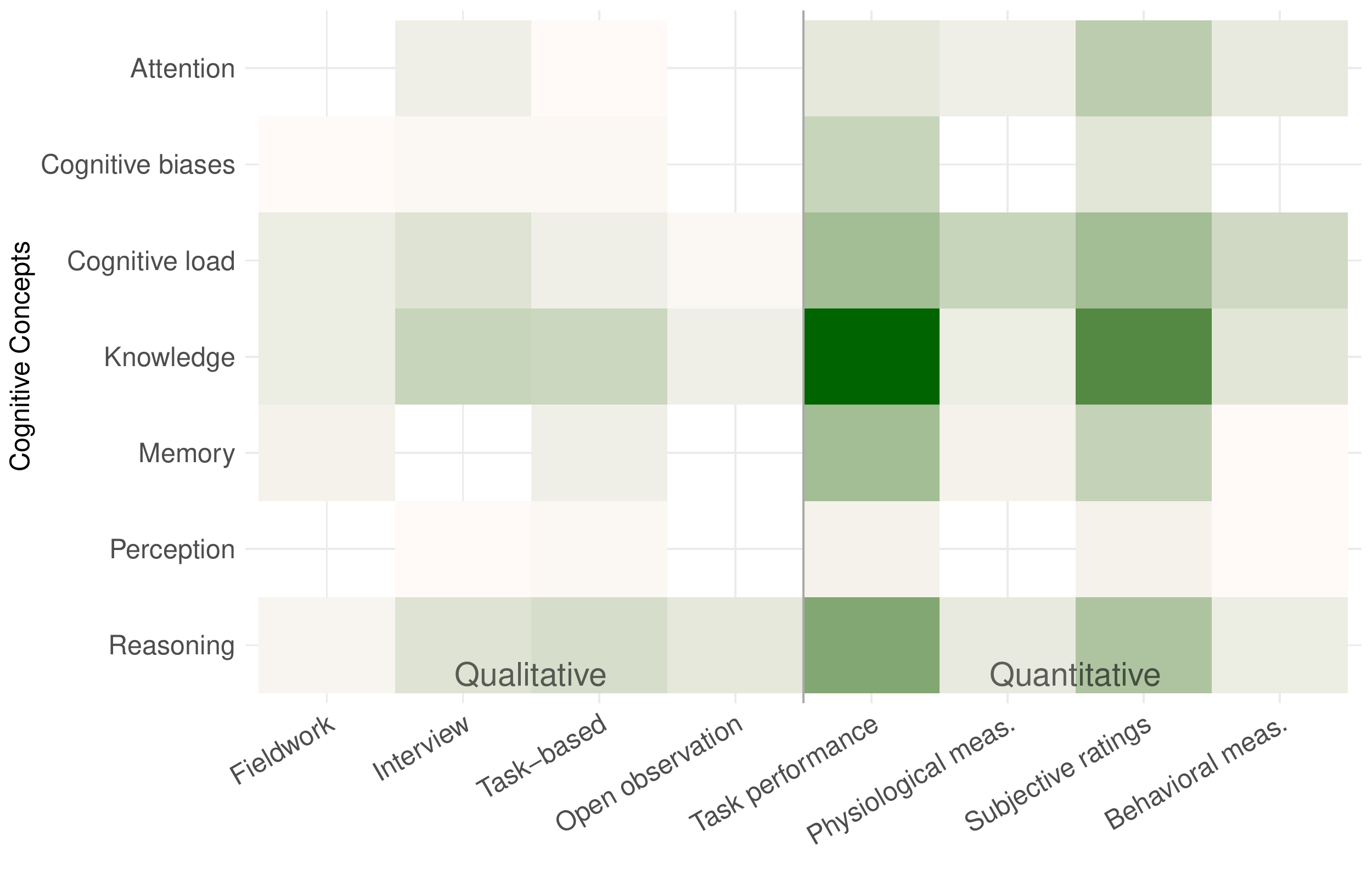}
  \caption{All publications.}
    \label{fig:taxonomy_method_all}
\end{subfigure}%
\begin{subfigure}{.32\textwidth}
  \centering
  \includegraphics[width=\linewidth]{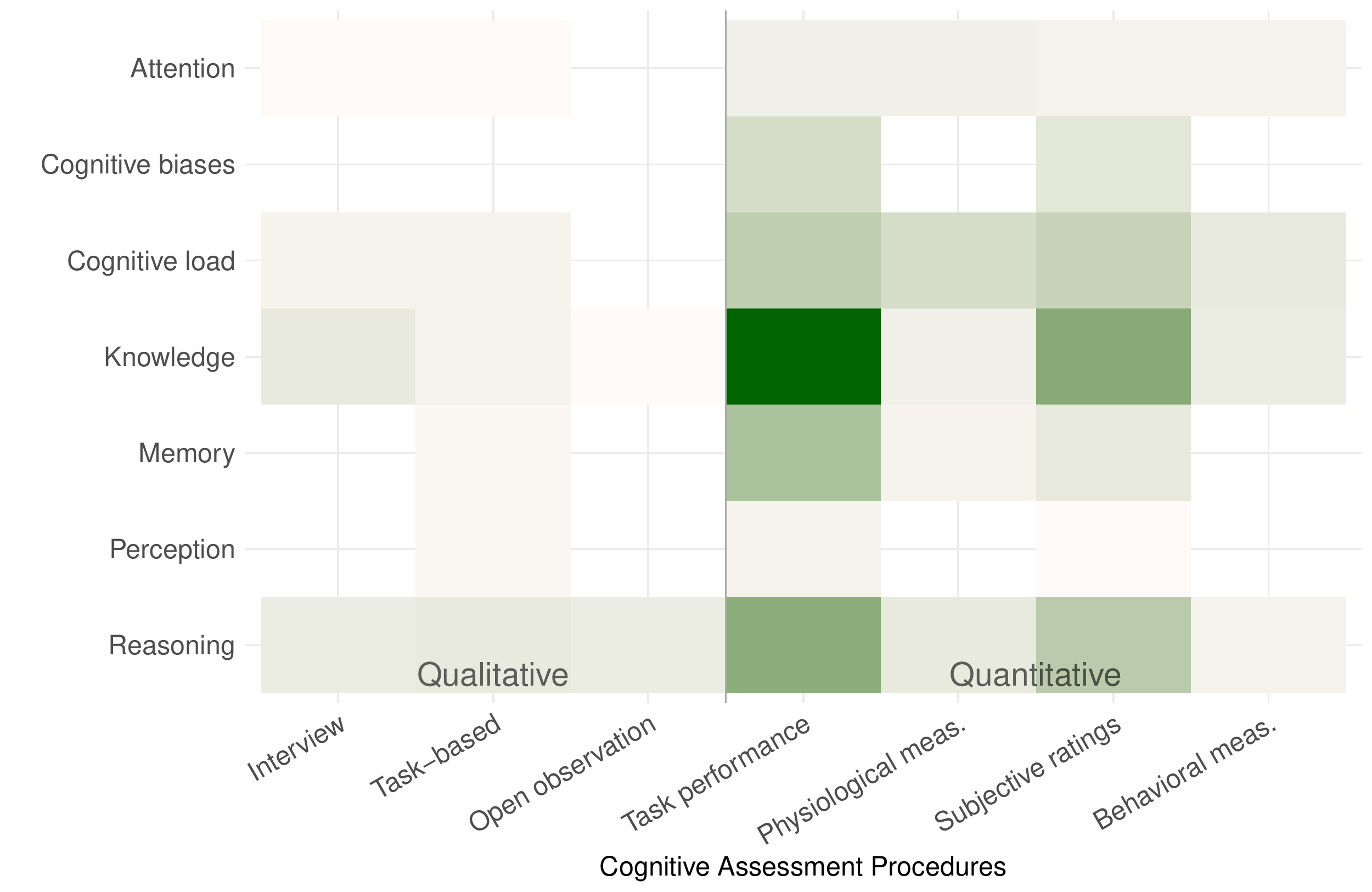}
  \caption{Publications in Academic context.}
  \label{fig:taxonomy_method_academia}
\end{subfigure}
\begin{subfigure}{.32\textwidth}
  \centering
  \includegraphics[width=\linewidth]{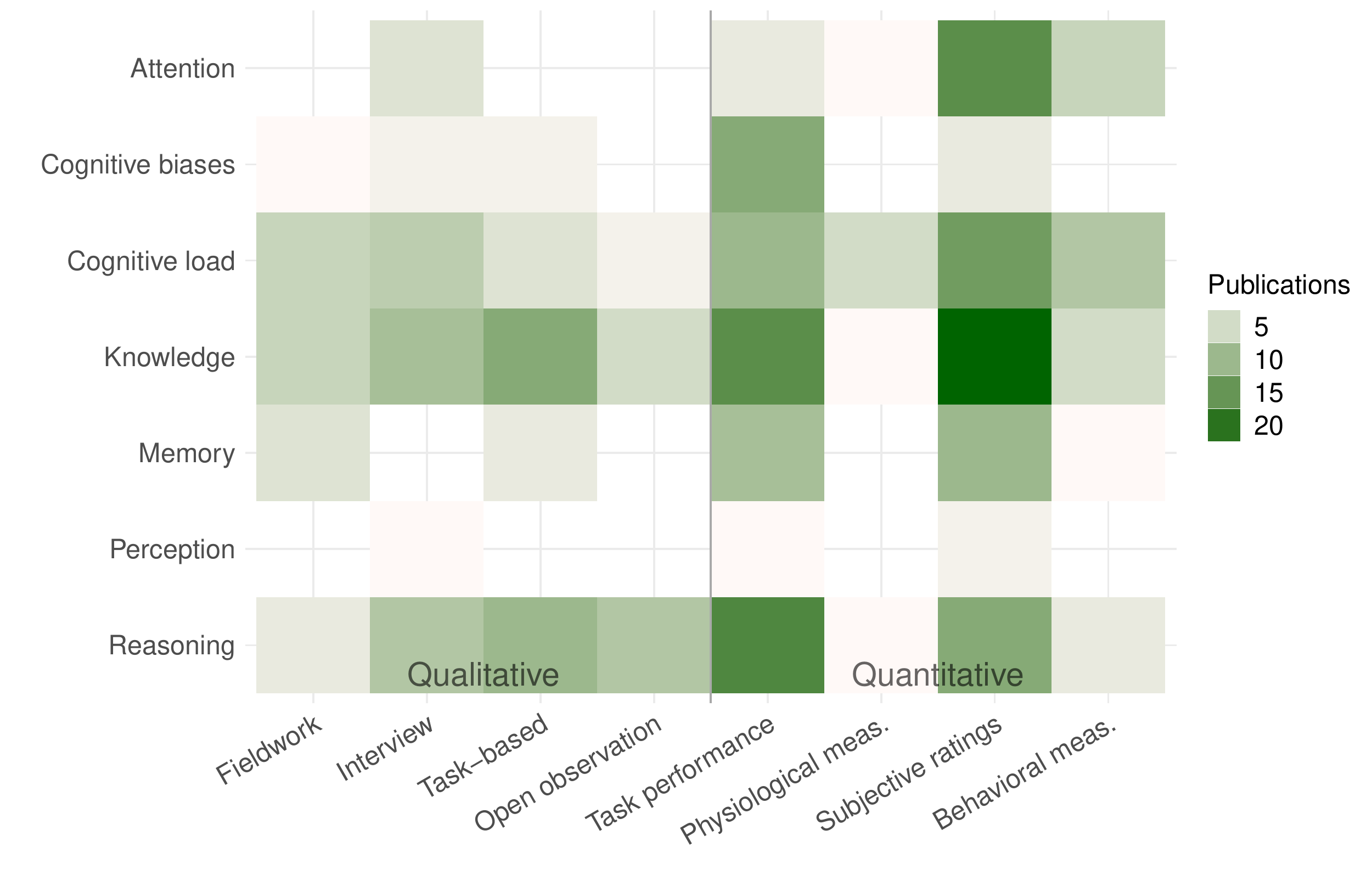}
  \caption{Publications in Industrial context.}
  \label{fig:taxonomy_method_industry}
\end{subfigure}
   \caption{Publications characterised by Cognitive Concepts according to different Assessment procedures.}
\label{fig:taxonomy_method}
\end{figure}

RQ2.4 poses the question of how cognitive concepts in SE have been assessed. Figure~\ref{fig:taxonomy_method} shows the number of primary studies covering the cognitive concepts according to different assessment procedures, both qualitative and quantitative. There are differences in how the concepts are assessed, and most assessments are quantitative. \emph{Knowledge}, \emph{Cognitive biases} (\emph{Errors}), \emph{Memory}, \emph{Reasoning}, and \emph{Attention} are mainly assessed using quantitative procedures, especially \emph{Task Performance} (e.g., a programming exercise) and \emph{Subjective Ratings} (e.g., self-reported understanding of requirements). \emph{Physiological} and \emph{Behavioural} procedures are under-utilised, perhaps due to limited availability of equipment for the former, and the lack of methodological training and guidance for both.

Qualitative assessments make use of \textit{Task-based} studies, \textit{Interviews}, and, to a lesser extent, \textit{Fieldwork} observations. Nonetheless, \textit{Fieldwork} is utilised for several concepts (e.g., \textit{Cognitive load}, \textit{Knowledge}, \emph{Reasoning}, and \textit{Memory}), whereas current research on cognition in SE lacks \textit{Open observations} assessments.

\begin{figure}
  \centering
  \includegraphics[width=.8\linewidth]{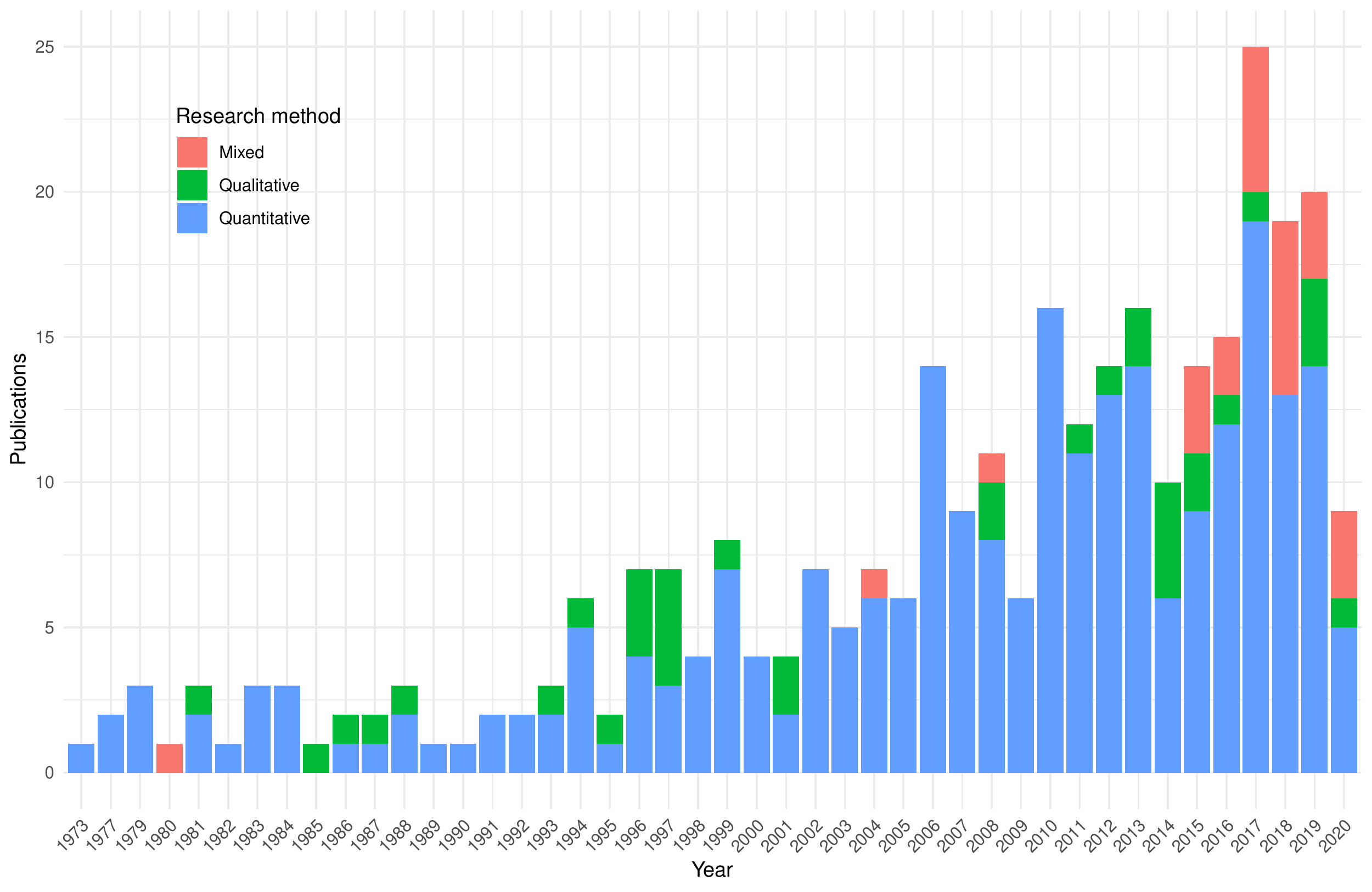}
  \caption{Research methods over time.}\label{fig:rm_years}
\end{figure}

The methods used likely reflect the methods familiar to the SE research community, the training typically included in researcher education in the field, and the traditions related to different methodological approaches. These aspects are all likely to change over time, and that is in fact what we found.
Figure~\ref{fig:rm_years} shows the number of publications in different years according to the type of research method.
Throughout the observed time period, quantitative methods represent the largest part of the research. However, qualitative methods are also utilised and, for some years, represent half of the total published research. More recently, mixed (or multimethod) approaches have become more common. These refer to studies which have two (or more) separate approaches side-by-side to investigate a topic, but they also include mixed-methods designs that strategically integrate quantitative and qualitative methods based on an explicit research philosophy, use specifically chosen mixing points, and explicate a logic for deriving valid results from the combination. The latter still appears to be less common than simpler combinations, and we have not attempted to separate the two in this analysis.

\begin{figure}[b]
\centering
  \includegraphics[width=.8\linewidth]{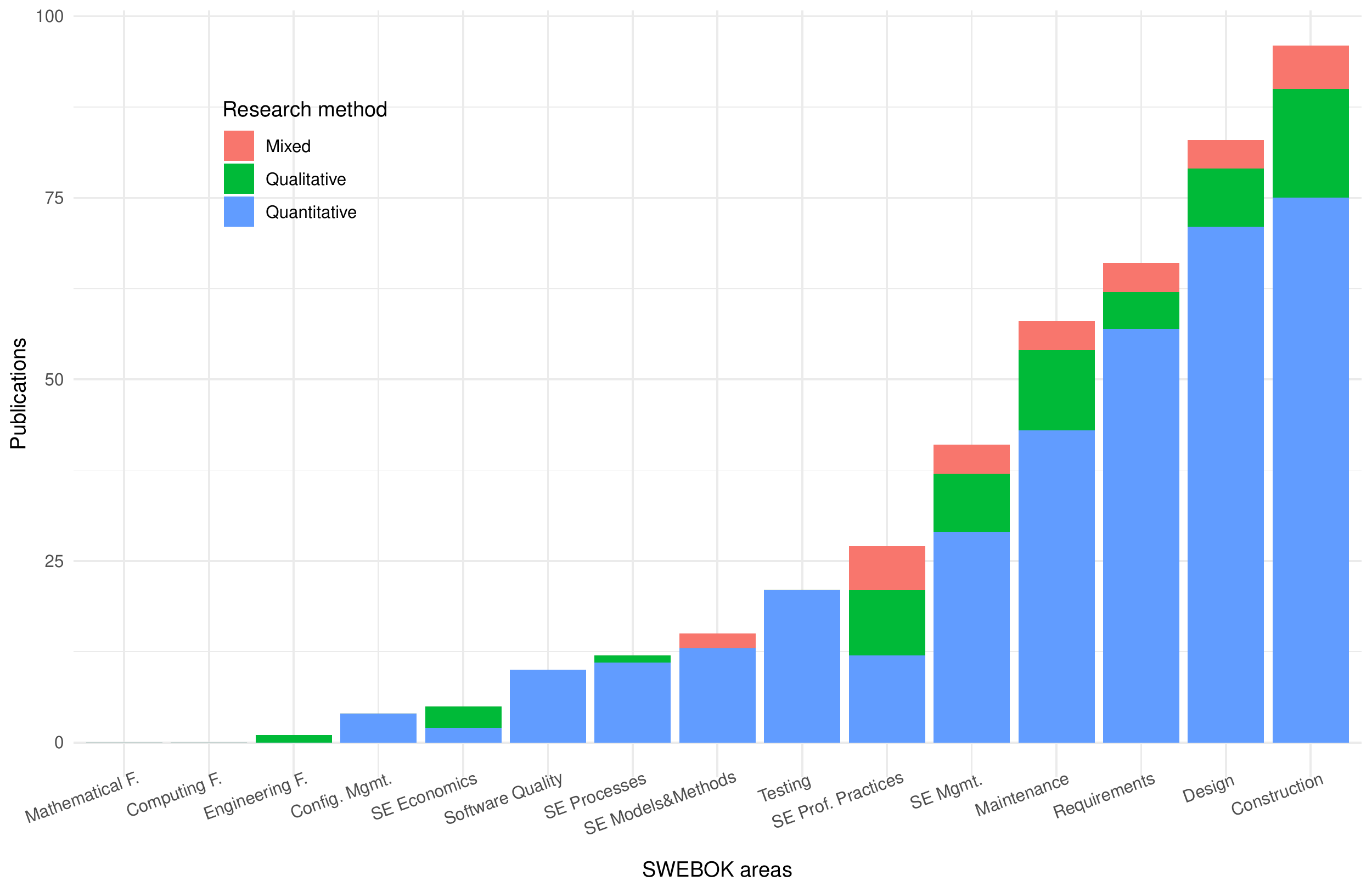}
  \caption{Research methods in SWEBOK areas.}
  \label{fig:swebok_researchmethods}
\end{figure}

Figure~\ref{fig:swebok_researchmethods} shows the distribution of publications covering different research methods in SWEBOK areas. Generally, the areas with most research are also the areas where qualitative and mixed methods are used the most. In SE Professional Practice, only half of the publications utilise a quantitative approach, and in SE Economics and Engineering Foundations, qualitative approaches outweigh the quantitative.

Finally, we examined how different cognitive concepts have been investigated using different assessment procedures. Among the qualitative approaches, \emph{Interviews} are used for almost all concepts, with \emph{Knowledge} and \emph{Cognitive load} being the two most common. Task-based procedures are also common, and most used for \emph{Problem-solving} and \emph{Knowledge}. \emph{Fieldwork} is used sparingly and mostly for \emph{Knowledge} and \emph{Cognitive load}, while \emph{Open observation} is also used sparingly and then mainly for \emph{Reasoning}, \emph{Problem-solving}, and \emph{Knowledge}.

The most common quantitative assessment procedures are \emph{Task performance} and \emph{Subjective ratings}. Both have a remarkably similar profile in terms of what concepts they have been used in. \emph{Knowledge}, \emph{Cognitive load}, and \emph{Problem-solving} are frequent targets of both. \emph{Task performance} has been used more for specifically assessing \emph{Intrinsic} and \emph{Extrinsic cognitive load} as well as \emph{Short-term memory}, but these are relatively infrequent uses. Among the other procedures, \emph{Physiological measures} have been used most frequently for \emph{Cognitive load} and to a slightly lesser extent for \emph{Problem-solving} and \emph{Knowledge}. \emph{Behavioural measures} are most commonly used to assess \emph{Cognitive load}, \emph{Knowledge}, and \emph{Attention}.

\begin{table}[t]
    \centering
    \caption{Examples of cognitive measures}
    \footnotesize
    \label{tab:cm_examples}
    \begin{tabular}{lp{11cm}}
        \toprule
        Assessment procedure & Cognitive measure \\
        \midrule
        \multicolumn{2}{c}{Qualitative} \\
         Fieldwork & Experiences concerning knowledge-transfer, task performance (Cognitive load/Knowledge); The occurrence of phenomena in a software development project that can be explained by a variety of theories (Reasoning/Cognitive biases) \\ 
         Interview & Experiences with job/project rotations (Knowledge); Merge conflict processes, barriers, and resolution strategies (Reasoning/Knowledge); Experiences concerning knowledge transfer (Knowledge)\\ 
         Task-based & Description of how experts and novices perform software design (Reasoning/Knowledge); Type and frequency of utterances, such as inquiries, questions and conjectures (Reasoning); Comparison of code comprehension by practitioners with theoretical model (Knowledge) \\ 
         Open observation & Shared mental models and teamwork characteristics (Reasoning/Knowledge); Identification and description of wasteful activities in software development (Cognitive load/Knowledge) \\ 
         \midrule
         \multicolumn{2}{c}{Quantitative} \\
         Task performance & Influence of anchoring on estimation performance (Cognitive biases); Balance of domain-specific and programming-specific information in the mental representations participants had of a program at different points in time (Knowledge); Ability to identify and fix bugs in the source code (Cognitive load/Knowledge) \\ 
         Physiological measures & Electrical signal emitted from muscle nerves (Reasoning/Knowledge); Pupillography (Cognitive load); Eye fixation (Reasoning) \\ 
         Subjective ratings & Visual memory capacity (Memory); Vulnerability of SE tasks to interruptions, dependent on suspension length and number of nested interruptions (Attention); Perceived cognitive load (Cognitive load) \\ 
         Behavioural measures & The proportional fixation time on each area of interest (Attention); Active time spent in IDE (Attention) \\ 
         \bottomrule
    \end{tabular}
\end{table}

We have also extracted cognitive measures, if identifiable, from the primary studies. Table~\ref{tab:cm_examples} illustrates a sample for each assessment procedure, indicating also which cognitive concept is measured. The complete list of measures is available in the online supplementary material~\cite{dataset}.

\subsection{Use and value of psychological theories}

We now turn to the role of theory in studies on cognition in SE, addressing RQ2.5. The examined papers mention a wide range of different theories, theoretical concepts, or conceptual frameworks as general inspiration. Some papers utilise theories or conceptual frameworks developed within the field of computing, such as the cognitive dimensions of notations framework~\citep{green1989}, ``physics'' of notations~\citep{moody2009}, and cognitive informatics~\citep{wang2003}. However, the vast majority of papers do not use theory or theory-grounded conceptual frameworks to situate the research, to inform their research design, or to explicitly draw hypotheses. No papers use their results to clearly and explicitly contribute back to theory, i.e., by refuting, supporting, altering, or extending a theory.

Studies dealing with \emph{Attention} use no psychological theories specific to attention. Most papers in this area appear in combination with other cognitive concepts, and many of the theories mentioned are related to cognitive load. Program comprehension is a frequent research topic among these papers. It is perhaps somewhat surprising that specific psychological theories of attention are not more specifically included as part of the theoretical palette in such studies.

Studies on \emph{Memory} mostly do not mention any specific memory theory, but some papers do mention theoretical concepts such as chunking (e.g.,~\cite{rosson1990}), aids for memory failures~\cite{parnin2012}, and multi-store memory theory~\cite{coulter1983}. However, with the exception of chunking and other program comprehension theories relating to memory, these are mainly part of setting the scene for the research and not explicit theories that would be used for informing the research design or hypotheses.

Papers examining \emph{Cognitive load} mostly mention some general form of cognitive load concept or theory, but usually without specifying in detail what the theory proposes or drawing any hypotheses specifically from the theory. \emph{Cognitive load} is also examined from the perspective of code or program comprehension, with theories such as bottom-up and top-down comprehension being mentioned in some papers (e.g.,~\citep{shneiderman1976}). Program comprehension studies have a long history during which theories specific to that field have been developed. This may in part explain why the theories are not among the ones found in cognitive psychology in general: the theories have already diverged from the more general theories in psychology. However, this cannot be said for most of the concepts examined in the papers.

There were no specific theories related to \emph{Perception} or \emph{Reasoning}. For the latter, some papers do mention broad theoretical ideas as background knowledge. Examples include problem space theory~\citep{kant1984}, the general idea that problem-solving involves searching a problem space, and Simon's theory on humans as information-processors that can be modelled similarly to computers~\cite{simon1979}. Similarly, \emph{Knowledge}, \emph{Cognitive control}, and \emph{Social cognition} papers did not include specific psychological theories.

While few of the papers make extensive use of theory, the cognitive concepts themselves play a central role in most of the papers. The specific cognitive concepts used in the vast majority of papers (which may be more detailed than the general categories in our taxonomy) really are central to the papers in their entirety, to the research approach, or to the arguments made. While they may not be well grounded in psychological theories, they do have specific intended meanings. For example, cognitive load is used in many papers to signify human limitations in information processing capacity. Several papers attempt to manipulate cognitive load in different ways to cause observable changes in task performance (e.g., error rates). Other papers use the cognitive concepts less explicitly: to provide a conceptual framework to explain what is studied, to inform the study design without providing an explicit theory, or to situate the research in the field of human factors in SE.

However, the use of cognitive concepts is somewhat imprecise in many papers, largely due to the lack of a connection to established definitions and theories in psychology and cognitive science.
As an example of a particular pattern, a paper may purport to examine mental effort---a dynamic characteristic of an individual human---but actually ends up assessing task difficulty---a characteristic of the task.
Furthermore, some papers seem disconnected from the cognitive concept they claim to examine.
Their use of the concept is more of a passing motivation resembling an inspiration or argumentative device to support the relevance of the study, or an assumption or conjecture on the authors' part.
An example of this pattern would be a paper arguing that a specific cognitive process is the underlying cause for specific human behaviour that the study has observed in software development, but does not actually demonstrate that this is the case.

Thus, the value of the cognitive concepts appears to differ a lot between the papers. Authors whose own conceptual development is thorough succeed better at producing trustworthy results. This applies both in papers that seek to deepen the understanding of cognition in SE and in papers that seek to introduce new methods or tools to aid cognitive processes among software engineers.

\section{Discussion}
\label{sec:discussion}

How software engineers think has been studied for over five decades, leading to insights on how they work and how individuals' performance can be improved through means of tooling, work organisation, and training.
Software construction has been in the centre of attention (98 primary studies), followed by software design (84 primary studies), and software requirements (66 primary studies).

We described the construction of a taxonomy for cognitive concepts and assessment procedures, and used it in a literature survey on research on cognition in SE. The taxonomy is based on expertise from cognitive psychology and related areas, and provides a viewpoint on research in SE that relates to cognitive processes. Our application of the taxonomy illustrates how it can organise and improve research by better utilising concepts from psychology and cognitive science as reference fields. Our taxonomy also shows the breadth of assessment options that exist and that researchers should consider.

We asked how cognitive concepts relevant for SE can be classified and measured (RQ1). Our taxonomy, and the examples of papers covering the cognitive concepts and assessment procedures, provides an answer to this question. We propose that cognition in SE should be conceptualised by drawing from reference fields such as psychology and cognitive science, where decades of research have resulted in well-defined and empirically defensible conceptualisations. Trying to invent new concepts to investigate cognition in SE will likely result in conceptual confusion and in a set of concepts which are not suitable to investigate cognition.

We investigated what is the state of the art in cognitive concepts in SE (RQ2). The most developed strands of research seem to be in the areas of the four life-cycle stages \emph{Software Requirements}, \emph{Design}, \emph{Construction}, and \emph{Maintenance}. Most research is quantitative and focuses on \emph{Knowledge}, \emph{Cognitive load}, \emph{Memory}, and \emph{Reasoning}. Surprisingly, there is a clear lack of research on cognitive concepts in \emph{Software Testing}. The state of the art overall appears fragmented when viewed from the perspective of cognition. There is a lack of use of cognitive concepts that would represent a coherent picture of the cognitive processes active in specific tasks. The research methods are focused mainly on approaches that are familiar to SE researchers and, while SE research seems to adopt new methods, the uptake of methods from fields like psychology seems slow. Nevertheless, among the questions and topics examined, there are good examples of research in cognition, as shown in this paper.

We asked what gaps and potential future directions exist in the field of cognition in SE (RQ3).
Our literature survey shows that while cognitive concepts have been studied in SE for more than 50 years, there is no single consensus on how to view or study cognition in this field.
The field is organised mainly in terms of the SE activities that practitioners perform rather than in terms of the cognitive processes that are active while they perform them.
Thus, the state of the art is specific to each area within SE and based on the traditions of that area.
The advantage is that researchers address problems that are understandable within a taxonomy of SE concepts (e.g., SWEBOK)---such as program comprehension during software construction.
The drawback is that the state of the art does not present an in-depth explanation of the cognitive processes that are active and that can lead to different outcomes in performing the activities.

In the following sections, we i) discuss the research gaps in each cognitive concept and provide recommendations for future research, ii) compare and industrial context and provide recommendations on the latter, iii) provide exemplary use cases for using the taxonomy.

\subsection{Gaps and future research directions}\label{sec:areas_discussion}
\textit{Perception} is the least studied concept.
This may be the case because it commonly deals with making sense of and organising visual information.
However, the predominant medium in software development is text, both in natural language and code. 

We showed that \textit{Perception} is investigated in the context of the SE tasks that make use of visuals (e.g., UI sketches in requirements engineering or architectural diagrams in software design).  
Nowadays, making sense of and communicating the large amount of data collected at every step of the development life-cycle necessitates visual abstractions, for which \textit{Perception} plays an important role.
In particular, the areas of \textit{Software Processes} and \textit{Software Quality} fit the study of \textit{Perception} due to their reliance on graphical information---e.g., notations used for process modelling or metrics visualisation used for software quality improvement.

Perception is inherently difficult to assess using unobtrusive techniques, such as open observations or fieldwork.
Accordingly, we did not find any publication investigating this concept using these procedures.
More intrusive approaches---e.g., physiological measures obtained using biometrics---can fit the study of \textit{Perception.}
For example, biometrics are widely used in the area of code comprehension (e.g.,~\cite{siegmund2017}) and can be adapted to investigate visual information in SE. 
However, biometrics are under-utilised for the study of \textit{Perception}.

\perspective{Future research directions for the \textit{Perception} concept}{
\begin{itemize}[leftmargin=5mm]
 \vspace{-6mm}
    \item[$\star$] Investigate \textit{Perception} in areas of SE where visuals are used to summarise large amounts of data.
   \item[$\star$] Two areas that can particularly benefit from the study of \textit{Perception} are \textit{Software Processes} and \textit{Software Quality}. 
   \item[$\star$] We recommend employing \textit{Physiological measures} for the study of this cognitive concept.
\end{itemize}
}
\\

\textit{Knowledge} is among the most studied concepts in the literature, and it covers several areas of SE. 
As \textit{Knowledge} deals with how individuals incorporate new information and experiences in their work, this concept should be inherently present in every software development activity. 
Nevertheless, we show that \textit{Knowledge} has not been investigated in large sub-areas, such as \textit{Software Quality} and \textit{Configuration Management}. 

Studies of \textit{Knowledge} in areas such as \textit{Software Design} focus on how developers are able to acquire and use new information obtained from models like UML~\cite{cruz2009assessing}. 
In other words, the knowledge acquired by a study subject is used as an outcome to evaluate an intervention (e.g., the use of new UML symbols). 
The study of \textit{Knowledge} is transferable to other areas. 
For example, in the area of \textit{Configuration Management}, one can study how a release recommendation system impacts the knowledge that different stakeholders have of the system to be deployed---e.g., in terms of requirements included in the release.  

We found that \textit{Knowledge} is assessed based on individuals performing tasks (i.e., \textit{Task performance} and \textit{Task-based}) as well as using their \textit{Subjective rating}---these procedures are a natural fit for this cognitive concept. 
Other assessments procedures, such as \textit{Fieldwork}, \textit{Open observations}, \textit{Physiological} and \textit{Behavioural measurements} can be used to infer an individual's \textit{Knowledge}. 
These assessments combined offer an alternative to approaches that necessitate the active involvement of the individual participating in a study, and can mitigate threats such as apprehension bias~\cite{Sed12}.

\perspective{Future research directions for the \textit{Knowledge} concept}{
\begin{itemize}[leftmargin=5mm]
 \vspace{-6mm}
    \item[$\star$] All areas of SE can benefit from studying the \textit{Knowledge} concept.
   \item[$\star$] We especially recommend measuring \textit{Knowledge}, as the outcome of an intervention, in the areas of \textit{Software Quality} and \textit{Configuration Management}.
   \item[$\star$] We recommend investigating assessment procedures that can measure \textit{Knowledge} without requiring explicit actions from the participants. 
\end{itemize}
}
\\

We showed that \textit{Reasoning} has also been intensively studied in SE. 
Similarly to Knowledge, \textit{Reasoning} can be investigated throughout different SE areas as it relates to decision making and problem solving.
Although \textit{Reasoning} is an implicit process (as opposed to Knowledge which is analytical in nature) it can be manipulated, and its outcome observed.
For example, LaToza et al.~\cite{latoza2020} compared developers who could employ their own decision making strategy for debugging a program to developers who had to follow a given one.  

\textit{Software quality} and \textit{Testing} are areas involving problem solving skills, whereas \textit{Configuration management} and \textit{Software processes} involve decision making.
Nevertheless, these areas are understudied from a \textit{Reasoning} perspective.

This cognitive concept is investigated using more direct assessment procedures, such as \textit{Task performance} and \textit{Subjective ratings}, which are proxies for the \textit{Reasoning} process. 
Similarly to \textit{Knowledge}, procedures that do not require an intervention on the individual are not broadly utilised. 

\perspective{Future research directions for the \textit{Reasoning} concept}{
\begin{itemize}[leftmargin=5mm]
 \vspace{-6mm}
    \item[$\star$] Although \textit{Reasoning} is hard to observe as a direct outcome of a study, it can be manipulated in several tasks across SE areas, and future studies should seek to develop such research designs.
   \item[$\star$] We recommend studying the effects of manipulating \textit{Reasoning} specifically in the areas of \textit{Software Quality}, \textit{Testing}, \textit{Configuration Management}, and \textit{Software processes}.
   \item[$\star$] We recommend using procedures which can make the \textit{Reasoning} process explicit and that do not require direct measures---e.g., \textit{Fieldwork} and \textit{Open observation} accompanied by think-aloud protocol and \textit{Interviews}.
\end{itemize}
}
\\

\textit{Attention} is being studied more and more in the last 20 years. 
As opposed to other concepts, such as \textit{Knowledge} and \textit{Reasoning}, its study may not fit process-related areas which are involved with cross-cutting concerns---e.g., \textit{SE Processes}, \textit{Software Quality}, and \textit{SE Models \& Methods}.
\textit{Attention} is observable over a short time span, which makes its study particularly suited for individual activities such as software \textit{Construction} and \textit{Maintenance}.
For example, \textit{Attention} plays a role in code comprehension where focusing on lexical similarities between code elements is important~\cite{fritz2014using}.
We showed that areas where individual attention is important for a task---e.g., \textit{Testing} and \textit{Configuration management}---need to be better investigated. 

In the literature, \textit{Attention} is generally assessed through quantitative procedures, such as \textit{Subjective rating}, \textit{Behavioural} and \textit{Physiological} performance.
Due to its transient nature, it is difficult to assess this concept in purely observational settings (i.e., using procedures such as \textit{Fieldwork} and \textit{Open observation}).
Yet, some procedures such as \textit{Interview} and \textit{Task-based} assessment can uncover the motivation behind \textit{Attention} deficit or prolonged attentional states~\citep{MF15}.

\perspective{Future research directions for the \textit{Attention} concept}{
\begin{itemize}[leftmargin=5mm]
 \vspace{-6mm}
    \item[$\star$] \textit{Attention} is a transient phenomenon best investigated in SE areas requiring an individual to manage their focus in detail-oriented tasks.
   \item[$\star$] We recommend studying the effects of different states of \textit{Attention} in the areas of \textit{Testing} and release planning as part of \textit{Configuration Management}. 
   \item[$\star$] We recommend studying the motivation for changes in \textit{Attention} using qualitative approaches, if possible.
\end{itemize}
}
\\

In contrast to \textit{Attention}, the study of \textit{Memory} in SE decreased in the last 20 years. 
\textit{Memory} has an inherent duplicity associated to it.
Short term memory fits SE areas that includes more transitory tasks, such as code comprehension (in the area of \textit{Construction}) and code reviews (in the area of \textit{Maintenance}).
Long-term memory fits tasks which unfold over a longer span of time, such us maintaining traceability links in the area of \textit{Configuration management}. 
While the study of \textit{Memory} has covered short-term memory---for example, Baum~\citep{baum2019} showed the relationship between working memory and code review effectiveness---we showed that long-term memory needs more investigation.

\textit{Memory} is a fundamental component in many complex tasks, and holds information and supports mental models that individual software developers rely on in their work. Similarly to \textit{Attention}, \textit{Memory} can be assessed using a wide range of quantitative procedures, but lacks consideration from a qualitative perspective---e.g., to identify the reason behind memory lapses, consequences of recalling incorrect information, or to understand the background for reasoning and decision-making.

\perspective{Future research directions for the \textit{Memory} concept}{
\begin{itemize}[leftmargin=5mm]
 \vspace{-6mm}
   \item[$\star$] We recommend studying \textit{Memory} in areas for which long-term \textit{Memory} is important (e.g., \textit{Software processes} and \textit{Software Models \& Methods})
   \item[$\star$] We recommend supporting quantitative procedures with qualitative ones, such as \textit{Interviews} and Cognitive Task Analysis procedures.
\end{itemize}
}
\\

\textit{Cognitive load} is one of the concepts which has constantly received attention from the SE research community starting from the 90s, particularly regarding code comprehension in the areas of software \textit{Construction} and \textit{Maintenance}. 
In both areas, developers use several strategies to understand the use of, for example, APIs or the behaviour of existing programs written by others (e.g., for code reviews). 
For instance, Siegmund~\cite{siegmund2017} showed that beacons (i.e., semantic cues) are used to locate relevant code, and are associated with specific activation patters in the brain.

Activities in the areas of \textit{Configuration management}, such as understanding, constructing, and maintaining traceability links, or in \textit{Software quality}, such as understanding and characterising defects, have been neglected. 
As \textit{Cognitive load} is difficult to observe, researchers used proxy measures with procedures involving active individual participation (e.g., \textit{Task performance} and \textit{Subjective ratings}). 
Other procedures aim to quantify \textit{Cognitive load} using \textit{Physiological} and \textit{Behavioural} measures. 
Although under-utilised, qualitative approaches are complementing the study of \textit{Cognitive load} to elicit indicators of cognitive overload.  

\perspective{Future research directions for the \textit{Cognitive load} concept}{
\begin{itemize}[leftmargin=5mm]
 \vspace{-6mm}
   \item[$\star$] We recommend studying \textit{Cognitive load} in areas other than the ones strictly related to software construction, such as \textit{Configuration management} and \textit{Software Quality}.
   \item[$\star$] We recommend studying the reasons and consequences related to, and experience of cognitive overload using qualitative procedures, such as \textit{Interviews}.
\end{itemize}
}
\\

The research interest on \textit{Cognitive biases} has been sustained over the last 20 years.
Similarly to other concepts, we showed that research on \textit{Cognitive biases} focused on the different phases of the development life-cycle (i.e., from \textit{Requirements} to \textit{Testing}).
For example, Salman~\cite{salman2018} investigated how confirmation bias takes place in test case design, whereas Mohanani~\citep{mohanani2018cognitive} examined how fixation bias impacts creativity of requirements. 
In general, studies of \textit{Cognitive biases} in these areas elicit a specific bias and measure its impact on a task relevant to that area. 
The remaining areas are understudied with the exception of \textit{SE management}.

There are large gaps in the assessment procedures for \textit{Cognitive biases}.
As biases usually result in sub-optimal decisions when performing a task, researchers used \textit{Task performance} and \textit{Subjective ratings} to measure them.
Yet, quantitative assessment based on \textit{Physiological} or \textit{Behavioural} measures are missing from the SE literature. 
Similarly to other concepts, it is inherently difficult to observe \textit{Cognitive biases} and pin-point their effects.
Nevertheless, qualitative procedures can support researchers in uncovering the causes of some cognitive biases, as well as evaluating interventions aimed at reducing their detrimental effects.  

\perspective{Future research directions for the \textit{Cognitive bias} concept}{
\begin{itemize}[leftmargin=5mm]
 \vspace{-6mm}
   \item[$\star$] We recommend studying \textit{Cognitive biases} in areas where decision-making has cross-cutting impact, such as \textit{Software requirements}, \textit{Software Design}, and \textit{Software processes}.
   \item[$\star$] We recommend applying the procedures currently used to investigate \textit{Cognitive biases} during the initial phases of the life-cycle to \textit{Maintenance}, which is currently understudied.  
   \item[$\star$] We recommend applying assessment procedures used for the study of other concepts, such as \textit{Physiological} and \textit{Behavioural} measures. Qualitative approaches should be investigated as well. 
\end{itemize}
}
\\

Our study focused on cognition as seen from a psychological angle, extending to social cognition that still situates cognitive processes in individuals. However, cognition can also be viewed differently, as distributed across individuals, embedded and situated in artefacts and environments~\citep{hutchins1995}. Although our result set includes a few papers which use distributed cognition theory as a backdrop, our study does not answer questions about research in SE using this conceptualisation. Future research could address this and contrast the taxonomy in this paper against new concepts found, potentially extending the taxonomy.

\subsection{Cognitive concepts studies in Industry vs. Academia}
Although both industry and academia have received similar attention in the literature, the study of cognitive concepts over the different areas of SE varies between these contexts. 
In both contexts, the focus is on the different phases of the development life-cycle, from \textit{Requirements} to \textit{Maintenance}---specifically on tasks related to software \textit{Construction}.
Notably, the investigation of \textit{Knowledge} in the area of software \textit{Construction} presents the most published studies in both academia and industry.

For selecting which area to focus on, context matters. 
Industrial studies are in fact predominant in areas of cross-cutting concerns for SE---from \textit{SE management} to \textit{SE economics}---which are difficult to attain in a purely academic context.
As pointed out in Section~\ref{sec:areas_discussion}, cognitive concepts that are investigated over an extensive time span (e.g., long-term \textit{Memory}) are better suited for SE areas akin to Industrial context (e.g., \textit{Software professional practice}).
Similar examples are \textit{Attention}, \textit{Cognitive biases}, and \textit{Cognitive load}---i.e., concepts studied mainly in an industrial context, and in SE areas representing cross-cutting concerns.
Exceptions are the areas of \textit{Configuration management} and \textit{SE Economics} which are understudied despite fitting industrial investigations.

\perspective{Future research directions in Industrial Context (SE areas)}
{
\begin{itemize}[leftmargin=5mm]
 \vspace{-6mm}
  \item[$\star$] Cross-cutting concerns should be investigated in industry studies, as the conditions for them are difficult to replicate in academic settings.
  \item[$\star$] To investigate cognitive concepts over long time-spans, researchers should consider SE areas that capture the complexities of the relevant industrial context.
  \item[$\star$] Sharply focused aspects within SE areas and clearly demarcated cognitive concepts (e.g., limited in time) can be investigated in academic settings, but care must be taken when attempting to transfer findings to industry.
  \item[$\star$] Researchers should prioritise studies of cognitive concepts in areas suited for industrial investigation, such as \textit{Configuration Management} and \textit{SE Economics}, which are currently understudied.
\end{itemize}
}
\\

The assessment measures observed in the literature reporting studies in academic context heavily favour quantitative approaches. 
\textit{Fieldwork}---an approach which requires  extensive presence of the researcher in the environment---is never utilised to carry out academic studies on cognitive concepts. 
In this context, researchers study \textit{Knowledge} and \textit{Reasoning} using qualitative approaches.
Conversely, qualitative approaches are never used to investigate \textit{Cognitive biases}.

In industrial studies, there is more balance between qualitative and quantitative approaches. 
Qualitative studies, such as \textit{Fieldwork} and \textit{Task-based} observations are widespread in this setting as they are not disruptive of the day-to-day activities, for example, practitioners working in a company. 
To some extent, this applies to quantitative assessments, too.
Low-disruption and non-intervention approaches, namely \textit{Task performance} (i.e., for a task which can be measured offline, such as defects found in a review) and \textit{Subjective ratings} (i.e., in the form of quick surveys or experience sampling~\cite{GLN21}), are preferred. 
More invasive procedures, which require ad-hoc devices, such as \textit{Physiological measures}, are still uncommon. 
However, less invasive technologies have shown great potential in enabling biofeedback-based procedures in industry~\cite{ZCM17}. 

\perspective{Future research directions in Industrial Context (Assessment procedures)}
{
\begin{itemize}[leftmargin=5mm]
 \vspace{-6mm}
   \item[$\star$] Qualitative procedures, currently used in industrial studies, need to be used more in academic studies as the latter present a context that can be an adequate yet cheaper approximation of the former~\cite{FJW18} (e.g., in terms of recruiting subjects).
   \item[$\star$] Qualitative and quantitative procedures should be chosen according to research objectives. We recommend the use of advanced mixed-methods research designs when investigating issues cutting across multiple SE areas, with potentially more than one cognitive concept, and where there are multiple research objectives (e.g., in terms of description, explanation, and theory-building).
   \item[$\star$] We recommend considering unobtrusive technologies to unlock the use of \textit{Physiological measures} for investigating  cognitive concepts such as \textit{Perception} and \textit{Cognitive biases} in industrial context.
\end{itemize}
}
\\

In general, we call for the adoption of new methods in cognition in SE research.
Adopting methods brings with it the need to understand them and their underlying theories, which can advance the state of the art in the field.
Physiological measurements, but also more advanced qualitative and mixed-methods approaches can make the research on cognition in SE more detailed, rich, and meaningful.
A broader repertoire of methods would allow investigating cognition from a multitude of perspectives, and could lead to development of novel methods or variants of methods within SE research itself.

\subsection{Uses for the taxonomy}
 We argue that despite decades of research, there are plenty of opportunities for fruitful work in the area.
 Our development and application of a taxonomy of cognitive concepts and assessment procedures shows that it is possible to start organising the SE field in terms of cognition.
 Researchers and educators can use the taxonomy as a support for the following tasks.

\paragraph{Filtering existing studies.}
Researchers can apply the taxonomy to the primary studies covered in our SMS to obtain a list of papers about one or more cognitive concept of interest.
This use case fits teams of SE researchers wanting to build background knowledge in an area of cognition and cross-disciplinary teams, who already have background knowledge in psychology, wanting to understand the research gaps relative to a cognitive concept applied to SE.

In a more pragmatic use case, SE methodology experts, wanting to investigate the area of cognition, can use the taxonomy to filter primary studies based on one of the assessment procedures to i) identify gaps in the literature to steer their research agenda, and ii) understand the nuances of applying a given procedure when investigating cognition.  
For educators, the two strategies above can be used to select reading material for a specialised course or module on cognition in SE, or for a general course or module on SE research methods.

\paragraph{Classifying new studies.}
This taxonomy represents a reusable classification scheme for researchers interested in performing secondary studies in the area of cognition in SE and close research areas where cognitive concepts play an important role, such as Human-Computer Interaction.  
SE Researchers can use the terms introduced in this taxonomy to better position their work in terms of cognitive concepts and assessment procedures. 
The taxonomy provides a set of keywords to be used together with the existing SE-specific ones (e.g., ACM CCS\footnote{\url{https://dl.acm.org/ccs}}).
Depending on how specific a study is in terms of the cognitive concepts it investigates, it can be placed in a higher (less specific) or lower (more specific) category. Furthermore, the relationships between cognitive concepts can also be used in classifying new studies. Many phenomena of interest in SE involve complex combinations of cognitive concepts. Studies investigating more than one concept can accordingly be classified into multiple categories in the taxonomy. Such patterns of relationships between cognitive concepts can further detail the focus of a specific study and be used to classify it.

\paragraph{Supporting researchers preparing primary studies in the area.}
The taxonomy can guide researchers in structuring papers so that important information on cognition is easy to find.
Explicit sections need to be dedicated to the description of cognitive concepts of interest, their role in the study, and how they connect to the SE-specific activities. 
Similarly, the procedures need to be explicitly connected to the cognitive process they aim to assess.
Finally, a paper needs to discuss the results in light of the covered cognitive concepts.
Our experiences with extracting information from primary studies has shown us that authors seldom connect back to the cognitive concepts they originally set out to investigate.

\section{Conclusion}
\label{sec:conclusion}

In this article, we provided a taxonomy, a literature survey, and a gap analysis of cognitive concepts in SE. 
The taxonomy comprises the top-level cognitive concepts of perception, attention, memory, cognitive load, reasoning, cognitive biases, knowledge, and social cognition as well as qualitative and quantitative cognitive assessment procedures.
The taxonomy provides a useful means to filtering existing studies, classifying new studies, and supporting researchers preparing studies in the area.

In the literature survey, we systematically analysed 311 primary studies published since 1973.
We investigate the state of the art of cognitive concepts in SE based on the developed taxonomy and the SE Body of Knowledge. 
We showed that the most developed strands of research seem to be in the areas of the four software life-cycle stages, namely requirements, design, construction, and maintenance. 
Surprisingly, there is a clear lack of research on cognitive concepts in software testing.
Most research is quantitative, which is more familiar to most software engineering researchers, and focuses on knowledge, cognitive load, memory, and reasoning.
The state of the art appears fragmented when viewed from the perspective of cognition.
There is a lack of use of cognitive concepts that would represent a coherent picture of the cognitive processes active in specific tasks. 

We highlighted future research directions for each cognitive concept in the taxonomy.
In the future, SE practice where cognition is more centrally considered could see tools that augment the skills of individual developers, techniques that help reduce bias and errors, and a workforce that is more aware of their own cognition as the basis for successful software design and development.
The taxonomy and research synthesis presented in this paper give an overview of important cognitive concepts and assessment approaches that can help both researchers and practitioners take steps towards such goals.

\bibliographystyle{ACM-Reference-Format}
\bibliography{bibliography}
\end{document}